\documentclass{article}

\usepackage{arxiv}

\usepackage[utf8]{inputenc} 
\usepackage[T1]{fontenc}    
\usepackage{natbib}
\usepackage{doi}
\usepackage{lipsum}
\usepackage{algorithm}
\usepackage{algpseudocode}
\usepackage{amsmath}
\usepackage{amssymb}
\usepackage{float}
\usepackage{cancel}
\usepackage{enumitem}
\usepackage{mathtools}
\usepackage{dsfont}
\setcitestyle{square}
\usepackage{hhline} 
\usepackage{cleveref}
\usepackage[inkscapelatex=false]{svg}
\usepackage{soul}
\usepackage{bm}

\usepackage{accents}
\usepackage{trimclip}
\newcommand{\halfapprox}{\clipbox{0em 0em 0em 0.225em}{$\approx$}}

\usepackage{enumitem}

\usepackage{booktabs} 
\usepackage{multirow} 
\usepackage{dcolumn} 
\usepackage{longtable} 
\usepackage{tabularx}
\usepackage{makecell}
\usepackage[flushleft]{threeparttable}
\usepackage[caption=false]{subfig}

\usepackage{enumitem}

\usepackage{acronym}
\usepackage{nomencl} 
\makenomenclature
\usepackage{framed}
\usepackage{comment}

\DeclareUnicodeCharacter{0398}{\ensuremath{\Theta}}
\DeclareUnicodeCharacter{0394}{$\Delta$}

\title{Comparative Analysis and Evaluation of Aging Forecasting Methods for Semiconductor Devices in Online Health Monitoring}

\author{
 Adrian Villalobos \\
  Electronics \& Computer Science Department\\
  Mondragon University\\
  Mondragon, 20500 (Spain) \\
  \texttt{adrian.villalobos@alumni.mondragon.edu} \\
   \And
  Iban Barrutia \\
  Electronics \& Computer Science Department\\
  Mondragon University\\
  Mondragon, 20500 (Spain) \\
  \texttt{ibarrutia@mondragon.edu} \\
  \And
  Rafael Peña-Alzola \\
  Electronics \& Electrical Engineering Department\\
  University of Strathclyde\\
  Glasgow, G11RD (UK) \\
  \texttt{rafael.pena-alzola@strath.ac.uk} \\
  \And
  Tomislav Dragicevic \\
  Wind \& Energy Systems Department\\
  Technical University of Denamrk (DTU)\\
  Lyngby, 2800 (Denmark) \\
  \texttt{tomdr@dtu.dk} \\
  \And
  Jose I. Aizpurua \\
  Computer Science and Artificial Intelligence Department, University of the Basque Country (UPV/EHU)\\
  Ikerbasque, Basque Foundation for Science\\
  San Sebastian, 20018 (Spain) \\
  \texttt{joxe.aizpurua@ehu.eus} \\
}

\begin{document}
\maketitle
\begin{abstract}
Semiconductor devices, especially MOSFETs (Metal-oxide-semiconductor field-effect transistor), are crucial in power electronics, but their reliability is affected by aging processes influenced by cycling and temperature. The primary aging mechanism in discrete semiconductors and power modules is the bond wire lift-off, caused by crack growth due to thermal fatigue. The process is empirically characterized by exponential growth and an abrupt end of life, making long-term aging forecasts challenging. This research presents a comprehensive comparative assessment of different forecasting methods for MOSFET failure forecasting applications. Classical tracking, statistical forecasting and Neural Network (NN) based forecasting models are implemented along with novel Temporal Fusion Transformers (TFTs). A comprehensive comparison is performed assessing their MOSFET ageing forecasting ability for different forecasting horizons. For short-term predictions, all algorithms result in acceptable results, with the best results produced by classical NN forecasting models at the expense of higher computations. For long-term forecasting, only the TFT is able to produce valid outcomes owing to the ability to integrate covariates from the expected future conditions. Additionally, TFT attention points identify key ageing turning points, which indicate new failure modes or accelerated ageing phases.
\end{abstract}


\section{Introduction}
The increased deployment of renewable energy plants has led to an increased research activity focused on prognostic models for semiconductor-based power modules as a way to reduce the operation and maintenance costs through predictive maintenance (\citealp{Semicon_24}). The main goal of failure prognostics is the reliable and accurate remaining useful life (RUL) forecast of the component under study by modelling the analysed failure mode and forecasting the future ageing evolution (\citealp{DePater_23}).

Semiconductors are used as electronic switches due to their high efficiency and fast switching properties (\citealp{zhang2023review}). However, they are often ranked as having the lowest reliability in power systems (\citealp{Novak_21}). The main failure modes of semiconductors can be classified into sudden and ageing failures (\citealp{Abu_19}). Sudden failures are caused by random phenomena such as cosmic radiation or electric discharge\textcolor{blue}{, e.g. (\citealp{Kang2025})}. In contrast, aging failures are caused by environmental or operational stress that exceeds a failure threshold limit, \textcolor{blue}{e.g. (\citealp{zubizarreta2025uncertainty})}. With regard to aging failures, the main focus of this research is on fatigue damage, which is caused by an imbalanced coefficient of thermal expansion of the different materials within the semiconductor structure. 

\begin{figure}[!h]
	\centering
	\includegraphics[width=0.5\columnwidth]{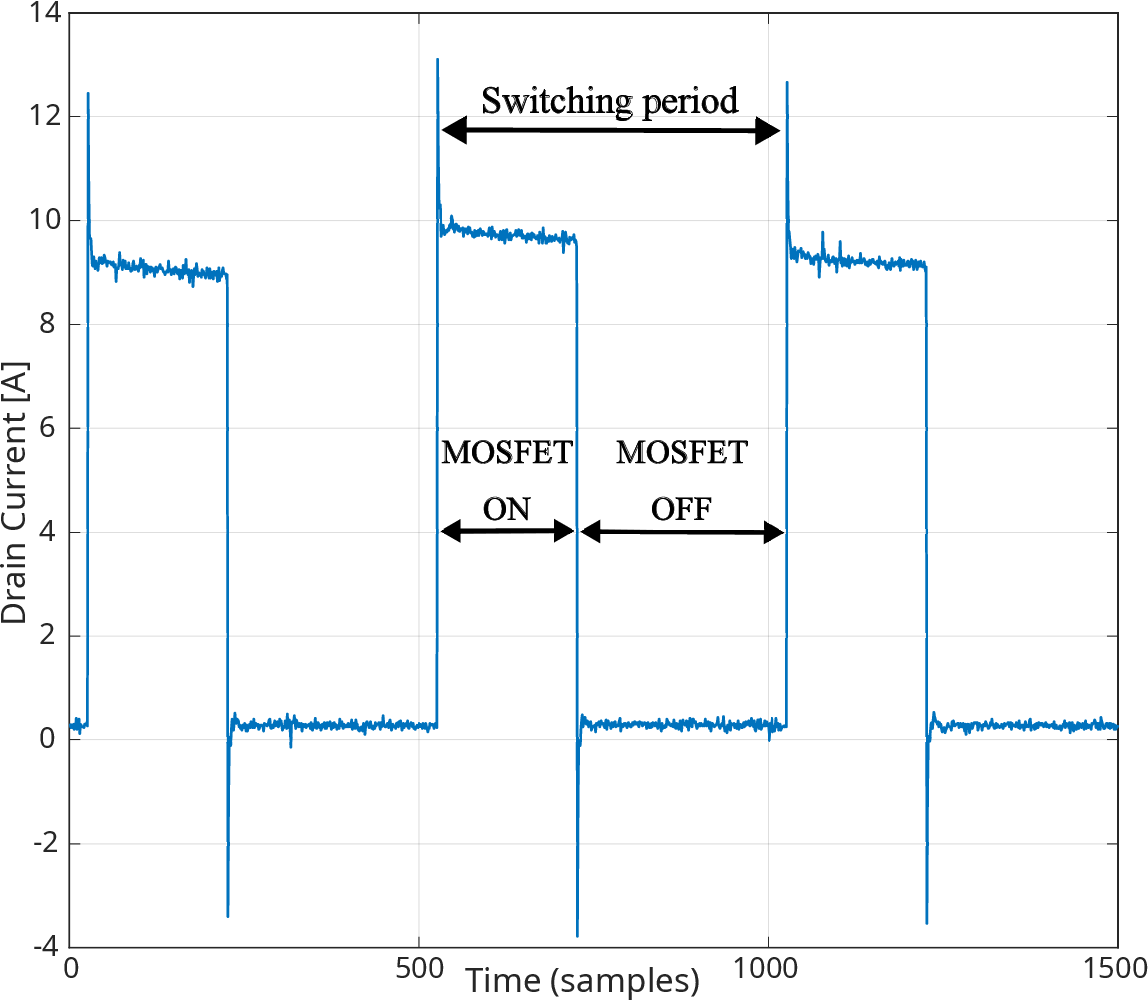}
	\caption{MOSFET degradation cycle example \textcolor{blue}{calculated from the case study dataset (cf. Section~\ref{sec:CaseStudy})}.}
	\label{fig:Cycling_Image_test36}
\end{figure}

MOSFETs are semiconductor devices used in different power applications (\citealp{lutz2018semiconductor}). The reliability and lifetime assessment of MOSFETs is complex and influenced by different aging processes that vary with cycling and temperature \textcolor{blue}{(\citealp{Fraser2025})}.  The main failure mechanism is die-attachment degradation, which leads to increased thermal impedance and higher device temperature, finally deriving in bond wire lift-off (\citealp{Celaya_12}). Thus, the degradation of lead-free solder die-attach results in an increase of the activation resistance or on-state, $R_{DS_{ON}}$, which depends on the junction temperature. Therefore, monitoring \textcolor{blue}{the evolution of} $R_{DS_{ON}}$ can be used to determine the RUL of the device under thermal stress. \textcolor{blue}{Namely, the MOSFET die attachment degradation cycle occurs due to the repeated operation of the MOSFET, which transits between the ON and OFF states and this causes an increase of the thermal impedance and device temperature, and eventually the occurrence of bond wire lift-off. Figure~\ref{fig:Cycling_Image_test36} shows an example of the MOSFET degradation cycle, estimated from the dataset used in the case study (cf. Section~\ref{sec:CaseStudy}). The drain current $I_D$ is used to estimate the $R_{DS_{ON}}$, which is the main precursor to predict the bond wire lift-off.}

\subsection{Related Works}

\textcolor{blue}{Artificial intelligence (AI) and machine learning (ML) applications are being widely developed for health monitoring applications for power electronic devices, e.g. design for reliability of power electronic systems (\citealp{Tomislav_19_AI}) or parameter design for converters (\citealp{Yanbo_24}). A comprehensive review of AI applications in power electronics is provided by (\citealp{Shuai_Rev_AI_21}). Among these applications, MOSFET prognostics have received significant attention, with NASA's MOSFET degradation dataset serving as a common benchmark (\citealp{Celaya_12}) -- see Section~\ref{sec:CaseStudy} for more details. Prognostics for MOSFETs rely on key failure precursors including $R_{DS_{ON}}$, drain to source voltage $V_{DS}$, and junction temperature $T_j$. Various methodologies, including statistical forecasting models, state-space approaches, and ML methods have been explored to enhance the accuracy of RUL prediction.}

\textcolor{blue}{Early studies used time-series and regression-based approaches to forecast degradation trends. (\citealp{UniVall_19}) analysed the deviation of $R_{DS_{ON}}$ from its pristine condition as a health indicator for RUL estimation. Under the assumption of an exponential degradation model, they explored Least Squares, Horizontal Average, and their linearized versions to improve one-step-ahead predictions. Similarly, (\citealp{Wang_22})  compared ARIMA and Gauss-Newton nonlinear regression, evaluating their accuracy in predicting $\Delta R_{DS_{ON}}$. Their results demonstrated that ARIMA provided more precise RUL estimates at different degradation stages.}

\textcolor{blue}{With the rise of data-driven methods, different ML models have demonstrated improvements in predictive accuracy. (\citealp{Li_2018}) integrated Echo State Networks (ESN) with Particle Filters (PF) to model the degradation process. The ESN parameters were dynamically updated using $R_{DS_{ON}}$ as a precursor. In an alternative approach, (\citealp{baharani2019real}) proposed a cloud-based monitoring framework using Long Short-Term Memory (LSTM) networks to track the variation of $R_{DS_{ON}}$, showing higher accuracy than Kalman and Particle Filters. Building on this, (\citealp{kang2024data}) developed an LSTM-based prognostic approach, trained and validated on NASA’s dataset with different training/testing proportions to optimize regression performance.}

\textcolor{blue}{Beyond deep learning, alternative ML approaches have emerged for enhanced feature extraction and predictive reliability. (\citealp{kahraman2024machine}) designed an ML pipeline integrating feature extraction, classification, and Bayesian Ridge Regression (BRR) for RUL prediction. RUL estimation was performed only if the classifier diagnosed the MOSFET in a pre-failure state. Feature-based methodologies have also proven effective in identifying MOSFET degradation patterns. (\citealp{Qian_22}) compared Nonlinear Auto-Regressive Model with Exogenous Inputs (NARX) and Partial Least Squares (PLS) regression, both combined with a Cumulative Sum (CUMSUM) analysis. Their approach demonstrated superior accuracy compared to Principal Component Analysis (PCA) and exponential weighted moving average (EWMA-PCA) in detecting early-stage degradation.}

\textcolor{blue}{Stochastic models offer another perspective on MOSFET prognostics by incorporating uncertainty in failure evolution. (\citealp{zhao2018health}) formulated MOSFET degradation as a Continuous-Time Markov Chain (CTMC), integrating it into Cox’s Proportional Hazards (PH) model to estimate time-to-failure. Here, filtering and discretizing $R_{DS_{ON}}$ measurements using \textit{k}-means clustering allowed for confidence-interval-based RUL predictions. Addressing the inherent variability in failure thresholds, (\citealp{wu2024remaining}) proposed a nonlinear Wiener process, continuously updating model parameters until reaching a probabilistic failure threshold, enabling a more adaptable prediction framework.}

\textcolor{blue}{Beyond MOSFETs, similar prognostics methods have been applied to Insulated Gate Bipolar Transistors (IGBTs), using different health indicators as failure precursors: collector-emitter voltage, $V_{ce}$, collector current, $I_{c}$, junction temperature, $T_j$, or transient thermal impedance, $TI$. These precursors capture degradation mechanisms related to gate oxide wear out, bond wire degradation, and solder fatigue, which are common aging failure modes in IGBTs (\citealp{zubizarreta2025uncertainty}).} \textcolor{blue}{Early approaches focused on time-series modeling and statistical estimation. (\citealp{Alireza_16}) proposed an RUL estimation framework based on Temporal Delayed Neural Networks (TDNNs). The model took four sequential measurements of $V_{ce}$ and their variations to estimate the component’s health state. A moving average filter was applied to remove noise, and a Normal distribution assumption helped refine predictions. This health-state was then combined with Maximum Likelihood Estimation (MLE) to infer confidence intervals for the RUL, offering a probabilistic failure prognostics.}

\textcolor{blue}{Unsupervised learning techniques have also been explored to capture degradation patterns without labeled failure data. (\citealp{Rigamont_18}) developed an ensemble of Self-Organizing Maps (SOMs) for component- and population-based degradation state identification. The ensemble dynamically weighted population-based SOMs providing general degradation trends with component-specific SOMs tailored to individual device behavior. Evaluations on an experimental dataset containing $V_{ce}$, $I_{c}$, and case temperature $T_{c}$ measurements showed reduced classification errors compared to static aggregation and single-component SOM models.}

\textcolor{blue}{Deep learning architectures have been also used to extract complex IGBT degradation features. (\citealp{Xiao_22}) introduced a self-attention-based Neural Network (NN) trained on a dataset containing $T_j$, $I_{c}$, $V_{ce}$, and package temperature. Instead of directly using raw sensor data, the study post-processed the signals to compute transient thermal impedance ($TI$), a direct failure precursor for IGBTs. 21 handcrafted statistical features were extracted and refined to form the dataset for improved prognostics. The model was evaluated under both offline and online training/testing strategies, demonstrating its adaptability for real-world applications.}

\textcolor{blue}{State-of-the-art graph-based and physics-informed ML approaches have recently gained traction in IGBT prognostics. (\citealp{Deng_2024}) introduced a Spatio-Temporal Fusion Graph Network (STFGN), where IGBT health states were modeled as graphs constructed from discrete voltage levels, enabling a structured approach to degradation prediction. Meanwhile, (\citealp{lu2023remaining}) applied Physics-Informed Neural Networks (PINNs) for RUL estimation, integrating domain knowledge into the Neural Network's loss function. This approach enforced monotonic degradation constraints and set initial health state values, ensuring physically meaningful predictions. Expanding on this direction, (\citealp{Fassi_Rev_24}) provided an extensive review of power converter maintenance strategies based on physics-informed ML methods.}

Table~\ref{table:sota_organization} displays the synthesis of related work organised according to semiconductor technology, ageing precursor, methods, prediction horizon, and whether the approach focuses on forecasting or regression models (F/R).

\begin{table}[htb]
\small
    \centering
 \begin{threeparttable}
    \caption{\textcolor{blue}{Summary of recent semiconductor ageing forecasting works.}}
        \begin{tabular}{m{10em} m{5em} m{6em} m{10em} m{9em} m{2em} }
             \toprule
             \textbf{Reference} & \textbf{Technology} & \textbf{Precursor} & \textbf{Methods} & \textbf{Horizon} & \textbf{F/R}  \\
           \midrule
           (\citealp{UniVall_19})  &  MOSFET & $R_{DS_{ON}}$ & LS variants & 1 step ahead & F \\ 
           (\citealp{Li_2018}) &  MOSFET & $R_{DS_{ON}}$ & ESN-PF  & 70\%-30\%, LOO & F, R \\ 
           (\citealp{Wang_22}) &  MOSFET & $R_{DS_{ON}}$ & ARIMA, Gauss-Newton  & $t_p$ = [100, 120, 140, 160, 180, 200] min & F \\ 
           (\citealp{Qian_22}) & MOSFET & $R_{DS_{ON}}$,$I_{D}$,$V_{DS}$  & NARX/PLS-CUMSUM & Anomaly Det. & - \\ 
           (\citealp{zhao2018health}) & MOSFET & $R_{DS_{ON}}$ & CTMC-PHs & Different times [min] & R \\ 
           \textcolor{blue}{(\citealp{kahraman2024machine})} & \textcolor{blue}{MOSFET} & \textcolor{blue}{$R_{DS_{ON}}$} & \textcolor{blue}{BRR} & \textcolor{blue}{Different instants from pre-failed state} & \textcolor{blue}{R} \\ 
           \textcolor{blue}{(\citealp{wu2024remaining})} & \textcolor{blue}{MOSFET} & \textcolor{blue}{$R_{DS_{ON}}$} & \textcolor{blue}{Wiener Process} & \textcolor{blue}{Continuous parameter update} & \textcolor{blue}{R} \\ 
           \textcolor{blue}{(\citealp{kang2024data})} & \textcolor{blue}{MOSFET} & \textcolor{blue}{$R_{DS_{ON}}$} & \textcolor{blue}{LSTM} & \textcolor{blue}{train/test proportions on test \#36} & R \\
           (\citealp{baharani2019real}) & MOSFET & $R_{DS_{ON}}$ & LSTM & 104 samples & R \\ 
           (\citealp{Alireza_16})  &  IGBT & $V_{ce}$ & TDNN & Long-term & R \\
           (\citealp{Rigamont_18})  &  IGBT & $V_{ce}$, $I_{col}$, $T$ & SOMs & Classification & R \\ 
            (\citealp{Xiao_22})  &  IGBT & $TI$ & Self-attention NN   & Train/test 30, 50, 70\%. LOO, 30\% & F, R \\ 
            \textcolor{blue}{(\citealp{Deng_2024})}  &  \textcolor{blue}{IGBT} & \textcolor{blue}{$V_{ce}$} & \textcolor{blue}{STFGN}   & \textcolor{blue}{weak supervised training} & \textcolor{blue}{R} \\ 
            \textcolor{blue}{(\citealp{lu2023remaining})} &  \textcolor{blue}{IGBT} & \textcolor{blue}{$V_{ce}$}  & \textcolor{blue}{PINN}   & \textcolor{blue}{Train/test 80\%/20\%, out-of-sample}  & \textcolor{blue}{R} \\
            \bottomrule           
        \end{tabular}
        \label{table:sota_organization}
\begin{tablenotes}[flushleft]
\setlength\labelsep{0pt}
\scriptsize
    \item[]\textbf{Legend}: F/R: forecast or regression; LOO: leave one out.
\end{tablenotes}
 \end{threeparttable}
\end{table}	

Table~\ref{table:sota_organization} \textcolor{blue}{displays} that the main failure precursor of MOSFETs is the activation resistance, $R_{DS_{ON}}$. Regarding the used predictive methods, mainly statistical methods (\citealp{UniVall_19,Wang_22,Qian_22,zhao2018health}), ML methods (\citealp{baharani2019real,kang2024data,kahraman2024machine,Rigamont_18,Alireza_16,Xiao_22, lu2023remaining, Deng_2024}), and combination of state-space and ML methods have been employed (\citealp{Li_2018}).

Most of the revised methods use regression models with covariates to inform the predictive model  (\citealp{Qian_22,zhao2018health,Alireza_16,baharani2019real,Rigamont_18, kahraman2024machine}), and only few of them focus on forecasting the future evolution of the ageing precursor using only past information. Regression models can predict the future ageing evolution, however, at the cost of monitoring additional variables and with limited ability to capture historical dependencies. In contrast, forecasting applications use only past information to predict the future aging evolution, which is more difficult, but also more cost-effective. Regression and forecasting methods mainly differ in time-dependency, incorporation of time-dependent properties such as autocorrelation or seasonality, and the treatment of covariates or exogenous variables.

Accordingly, the main focus of this work is to forecast the future evolution of the MOSFET aging trajectory based solely on past information. That is, this process requires learning historical degradation dynamics and forecasting future aging evolution, which is a challenging prediction task. Among the forecasting methods, most of the methods rely on one-step-ahead prediction models, which are sequentially integrated to predict longer horizons (\citealp{UniVall_19,Li_2018,Wang_22}). Interestingly, (\citealp{Xiao_22}) presented an IGBT health index forecasting through feature extraction, model learning and RUL prediction. \textcolor{blue}{Different from regression models, forecasting models may present a cost-effective monitoring strategy, which may be on-boarded in real time applications due to the ability of making predictions based on a single signal.}

\subsection{\textcolor{blue}{Contribution, Objectives \& Impact}}

Due to the rapid and widespread development of ML models for forecasting applications (\citealp{Benidis_22,Lim_21B}), the use of powerful new methodologies, such as transformer-based forecasting methods (\citealp{NIPS2017_Vaswani}), can generate useful insights for extended observations (\citealp{gilpin2023largescale}). They have been proven to be valid, especially for long-term predictions (\citealp{Biggio_23}). \textcolor{blue}{Unlike regression methods that consider multiple independent variables sampled at the same time instant of interest to explain the response variable, transformer-based forecasting models can incorporate covariates and historical dependencies.} 

Consequently, the main contribution of this work is the detailed comparative evaluation of classical forecasting methods with transformer-based forecasting methods, which have the ability to incorporate expected future covariates along with other influencing factors. To this end, first, it is necessary to adapt transformer models for MOSFET aging forecasting purposes through a fit-for-purpose architecture which uses forecasting covariates. Subsequently the capability of transformer models for short- and long-term forecasting has been exhaustively assessed and compared with a comprehensive benchmarking approach that includes different statistical, state-space based filtering, and ML methods.

There are works that analyze \textcolor{blue}{AI and ML} methods for power electronic applications (\citealp{Shuai_Rev_AI_21}). However, their focus is on the bibliometric research assessment and \textcolor{blue}{the} identification of research gaps.  \textcolor{blue}{In contrast, the goal of this research is the practical evaluation and critical comparison among failure forecasting methods for short- and long-term forecasting horizons through the implementation of classical statistical, state-space filtering and ML methods, along with state-of-the-art transformer-architecture based forecasting methods}. The proposed comparative evaluation has a direct impact on the health management of semiconductors in general and MOSFETs in particular. On the one hand, the proposed evaluation framework identifies trade-offs among state-of-the-art forecasting methods for MOSFET failure forecasting. On the other hand, transformer-based architectures are adapted for MOSFET failure forecasting, which can greatly improve long-term MOSFET failure forecasting activities.

\subsection{Organization}

The remainder of this article is organized as follows. Section~\ref{sec:LifetimeModelling_Review} reviews MOSFET ageing basics. Section~\ref{sec:Approach} presents the proposed ageing forecasting assessment approach. Section~\ref{sec:CaseStudy} defines the case study. Section \ref{sec:NumericalExperiments} applies the proposed approach to the case study. Section \ref{sec:Discussion} provides a discussion of the presented methodology, and finally, Section~\ref{sec:Conclusions} concludes.

\section{MOSFET Degradation Modelling}
\label{sec:LifetimeModelling_Review}

MOSFETs work in the ohmic region when ON, and in the cut-off region when OFF. The current through the semiconductor, $I_{DS}$, subject to drain to the source voltage, $V_{DS}$, is controlled by a voltage applied between the gate and the source terminals. Even though the MOSFET gate is electrically isolated, the gate driver needs to supply sufficient current to charge all the capacitances across the device. In particular, the capacitance between the gate and the drain (also called reverse capacitance) increases the effective input capacitance seen at the gate by orders of magnitude due to the Miller effect. Once the MOSFET is fully turned on, the voltage drops across source and drain can be modelled as an on-resistance, $R_{DS_{ON}}$, whose value is dependent on the junction temperature. 

The $R_{DS_{ON}}$ can be monitored by recording the values of the drain current $I_D$ and the drain to source voltage $V_{DS}$ in the device under ageing circumstances. On the other hand, monitoring the junction temperature is difficult and subject to inaccuracies (\citealp{Bossche}). Consequently, the main focus of this research is to track and forecast the evolution of $R_{DS_{ON}}$.

MOSFET end-of-life threshold criteria is not a deterministic value. Variations of $R_{DS_{ON}}$ are monitored, $\Delta R_{DS_{ON}}$, and it is generally assumed that a threshold value of $\Delta R_{DS_{ON}}$ 5\% would lead to failure of the MOSFET. Without assuming a fixed failure threshold, the main objective of this research is to forecast the evolution of $R_{DS_{ON}}$, which can be used to monitor the health of the MOSFET and predict the RUL, given a predefined failure threshold.

\subsection{State-space modelling}
\label{ss:state_space}

For the following, the on-resistance of the discrete MOSFET devices is considered under a pristine condition $R_{init}$; this value varies according to manufacturing tolerances. The temperature is assumed to be constant and equal to the value recommended by the manufacturer; this eliminates the temperature effects in the experiments.

The MOSFET ageing process results in an increment in the on-resistance with respect to the pristine condition, $\Delta R_{DS_{ON}}$, as a result of the degradation process due to the die-attach failure. An empirical degradation model for $R_{DS_{ON}}$ corresponds to exponential growth as a function of the time (\citealp{UniVall_19}):

\begin{equation}
\label{eq:ss2}
\Delta R_{DS_{ON}}(t)=\alpha (e^{\beta t} -1) 
\end{equation}

\noindent where $\alpha$ and $\beta$ are model parameters that can be static (estimated using e.g. least-squares) or estimated on-line.

The degradation model  in  Eq. (\ref{eq:ss2}) can be converted into a dynamic model by considering the state-space representation. Assuming $\alpha$ and $\beta$ as constant parameters and calculating the derivative of $\Delta R_{DS_{ON}}$ results in the following:

\begin{equation}
\label{eq:ss2_der2}
\dot{R}_{DS_{ON}}(k+1)=R_{DS_{ON}}(k)(1+\beta)+\alpha\beta
\end{equation}

Equation (\ref{eq:ss2_der2}) can be discretized and written in canonical form. Alternatively, if resistance under pristine conditions is considered, $R_{init}$, the state-space form results in the following (\citealp{Dusmez_16}):

\begin{equation}
\label{eq:ss1}
R_{DS_{ON}}(t)=\alpha e^{\beta t} + R_{init}
\end{equation}

\begin{equation}
\label{eq:ss1_der2}
R_{DS_{ON}}(k+1)=R_{DS_{ON}}(k)(1+\beta)-R_{init}\beta
\end{equation}

\section{MOSFET Ageing Forecasting Models}
\label{sec:Approach}

In order to forecast the future failure evolution of the MOSFET for different prediction horizons, different modelling approaches have been designed and tested: (i) classical state-space tracking algorithms based on Kalman filter variants, (ii) classical statistical forecasting methods, (iii) NN-based forecasting methods, and finally (iv) transformer-based forecasting methods.

\subsection{Non-linear Kalman Filters}

The Kalman filter can be used to predict the next step in linear processes, subject to process and measurement uncertainty. However, the state-space models (cf. Section~\ref{ss:state_space}) describe a non-linear process, and  accordingly, non-linear variants of the KF are implemented. The \textcolor{blue}{Extended Kalman Filter} (EKF) linearised the state-space equations around the operation point. The \textcolor{blue}{Unscented Kalman Filter} (UKF) uses transformations to capture the propagation of the statistical properties of state estimates through the nonlinear equations.
Thus, EKF and UKF can capture the effects of nonlinearities in the prediction of the aging process.

\subsubsection{Extended Kalman Filtering (EKF)}

The Extended Kalman Filter (EKF) (\citealp{chui2017kalman}) considers the estimation of the states $x\in\Re^n$ of discrete-time controlled processes that is non-linear. The EKF is a KF that linearizes the system equations around the mean using simple Taylor series with partial derivatives. The equations of the nonlinear stochastic difference equation for the state vector $x\in\Re^n$ are:
	
	\begin{equation}
	x_{k} = f(x_{k-1}, u_{k-1}, w_{k-1})
	\end{equation}
	
	\noindent with the measurement $z\in\Re^n$ is:
	
	\begin{equation}
		z_{k} = h(x_k, v_k)
	\end{equation}
	
	\noindent the random variable $w_k$ and $v_k$ are the process and measurement noise, respectively ($w_k$=0.002 and $v_k$=0.01, in this work).
	\medskip
	
The estimation process for the non-linear systems begins with the linearisation of the system equations around the current operation point:
	
	\begin{equation}
    \begin{split}
		x_{k} \approx \tilde{x}_{k} + A(x_{k-1} -\hat{x}_{k-1} ) + W w_{k-1}  \\
	  z_{k} \approx \tilde{z}_{k} + H(x_{k} -\tilde{x}_{k} ) + V v_k
    \end{split}
	\end{equation}
	
\noindent with $x_{k}$ and $z_k$ the actual state and the measurements and $\tilde{x}_{k}$ and $\tilde{z}_{k}$ the approximate state and measurement vector, respectively. The variable $\hat{x}_k$ is the \textit{a-posteriori} estimation of the state at step $k$. The approximate state and measurement vectors are calculated as follows:
 
    \begin{equation}
    \begin{split}
		\tilde{x}_{k} = f(\hat{x}_{k-1}, u_{k-1},0)  \\
	  \tilde{z}_{k} = h(\tilde{x}_k, 0)
    \end{split}
	\end{equation}
 
The matrices $\bf{A}$ and $\bf{W}$ result from applying the Jacobian to the nonlinear functions:
	
	\begin{equation}
	\begin{split}
		A_{[i,j]} = \frac{\partial{f_{[i]}}}{\partial{x_{[j]}} } (\hat{x}_{k-1},u_{k}, 0) \\
		W_{[i,j]} = \frac{\partial{f_{[i]}}}{\partial{w_{[j]}} } (\hat{x}_{k-1},u_{k}, 0) \\
		H_{[i,j]} = \frac{\partial{h_{[i]}}}{\partial{x_{[j]}} } (\tilde{x}_{k}, 0) \\
		V_{[i,j]} = \frac{\partial{h_{[i]}}}{\partial{v_{[j]}} } (\tilde{x}_{k},  0) \\
	\end{split}
	\end{equation}
For simplicity in the notation, the subscript $k$ in the Jacobians was omitted. The prediction error is defined as:
	
	\begin{equation}
	\tilde{e}_{x_k} = x_k- \tilde{x}_k
	\end{equation}
	
\noindent and the measurement residual:
	
	\begin{equation}
		\tilde{e}_{z_k} = z_k- \tilde{z}_k
	\end{equation}
	 
	The equations for the processes can be written as follows:
	 
	\begin{equation}
 	\begin{split}
		\tilde{e}_{x_k} = A(x_{k-1} -\hat{x}_{k-1} ) + \epsilon_k \\
		\tilde{e}_{z_k} = H(x_{k-1} -\tilde{x}_{k-1} ) +\eta_k
  	\end{split}
	\end{equation}
	
\noindent with $\epsilon$ and $\eta$ are random variables with zero mean and covariance matrices $WQW^T$ and $VRV^T$; $Q$ and $R$ are covariance matrices of the noise and measurement noise with $p(w) \sim N(0,Q)$ and $p(v) \sim N(0,R)$, respectively. These equations resemble the differential and measurement equations of the discrete linear Kalman filter. Using superscript minus notation for the a priori estimation and in this occasion maintaining the discrete time subscript $k$ for the Jacobians, the complete set of the EKF equations starts with the initialisation process with the initial estimate for $\hat{x}_{k}^{-}=E[x_k]$ and $P^{-}_k=E[(x_k-\hat{x}_k)(x_k-\hat{x}_k)^T]$. The time update process consists of the following processes:

\begin{enumerate}
	
	\item Project the state ahead: 
    \begin{equation}
    \hat{x}_{k} = f( \hat{x}_{k-1},u_{k-1},0 )
    \end{equation}
	
	\item Project the error covariance ahead: \begin{equation}
    P^{-}_{k} = A_k P_{k-1} A_k^T + W_k Q_{k-1} W^T_k
    \end{equation}

\end{enumerate}

The measurement update consists of the following steps:

\begin{enumerate}
	
	\item Compute the Kalman gain:
    \begin{equation}
    K_k = P_k^{-} H_k^T(H_k P_k ^{-} H_k^T + V_k R_k V_k^T)^{-1}
    \end{equation}
	
	\item Update estimates with measurements $z_k$: 
    \begin{equation}
    \hat{x}_k=\hat{x}_k^{-}+K_k(z_k-h(\hat{x}_k^{-},0))
    \end{equation}
	
	\item Update the covariance error:
    \begin{equation}
    P_k=(I-K_k H_k)P^{-}_k
    \end{equation}
	
\end{enumerate}

The procedure is applied to the preceding equations that model the MOSFET degradation process Eqs. (\ref{eq:ss2_der2})-(\ref{eq:ss1_der2}) in order to estimate the values of the describing parameters $\alpha$ and $\beta$.

\subsubsection{Unscented Kalman Filtering (UKF)}

The UKF is centred around the idea to describe the state $x$ as a distribution defined by a small number of characteristic sampling points, named sigma points, that represent the Gaussian underlying distribution. The UKF operates in an iterative way through sigma point determination, state-estimation, and update stages.

\subsubsection*{Sigma point determination}

The state-estimation, $x_k$ is approximated by the  by the sigma points, which are calculated from the mean, $\overline{x}_k$, and covariance, $P_k$ of the previous step. For each iteration, $k$, the sigma points for the state variables are estimated as follows (\citealp{UKF_01}):

\begin{equation} 
\label{eq:UKF}
\begin{split}
x[0] & = \overline{x}_k \\
x[i]  & = \overline{x}_k+(\sqrt{(n+\lambda)P_{k}})_i; \text{for i=1,\ldots,n}\\
x[i] &= \overline{x}_k-(\sqrt{(n+\lambda)P_{k}})_{i-n}; \text{for i=n+1,\ldots,2n}
\end{split}
\end{equation}

\noindent where $\lambda$ is a scaling parameter, $\lambda=\alpha^2(n+k)-n$, and $\alpha$, $\beta$ and $K$ are all tunable parameters to determine the spread of the sigma points.

These sigma points are then fed into the state equation to generate a set of new sigma points for the state variables in the current step. After calculation of sigma points, they are passed through the state-estimation function in Eq. (\ref{eq:ss2}), and their weights are calculated. Finally, they are used to form a state prediction together with a covariance and cross-correlation matrix.

\subsubsection*{Update stage}

In the update step, the predictions are combined with measurements [cf. Eq.~(\ref{eq:ss2_der2})] to form the new estimate. A detailed description of the UKF can be found in (\citealp{UKF_01}).

\subsection{Classical Statistical Forecasting}

\subsubsection{Autoregressive Integrated Moving Average (ARIMA)}

ARIMA combines differencing with autoregression and moving average. The model can be written as follows (\citealp{Hyndman_Book}):
\begin{equation}
    \label{eq:ARIMA}
    \hat{y}_t = c + \phi_1 y_{t-1}^{\prime}+\ldots+\phi_p y_{t-p}^{\prime}+ \theta_1 \epsilon_{t-1}+\ldots+\theta_q \epsilon_{t-q} + \epsilon_t
\end{equation}

\noindent where $\hat{y}_t$ is the predicted value of the time-series. The precursors of the time series include lagged values of $y_t$ and lagged errors. The constants to estimate include the $p$ order of the autoregressive part, the $d$ degree of the first differencing involved, and the $q$ order of the moving average part.
\smallskip

These parameters have been adjusted with the Hyndman-Khandakar algorithm. This algorithm combines unit root tests, minimization of the Akaike Information Criteria (AIC) and maximum likelihood estimation to obtain an ARIMA model. The implementation of the ARIMA model is developed using the \texttt{forecasting} package of R (\citealp{Hyndman_24}).

\subsubsection{Holt's Linear Trend Method}

Applies exponential smoothing to univariate data, with trend and no seasonal pattern. \textcolor{blue}{Holt's linear trend method} can be described with the following equations (\citealp{Hyndman_Book}):

\begin{equation}
    \label{eq:forecast_ses}
    \hat{y}_{t+h|t}=l_t+hb_t
\end{equation}
\begin{equation}
    \label{eq:smoothing}
    l_t=\alpha^* y_t + (1-\alpha^*)(l_{t-1}+b_{t-1})
\end{equation}
\begin{equation}
    \label{eq:smoothing2}
    b_t=\beta^* (l_t-l_{t-1}) + (1-\beta^*)b_{t-1}
\end{equation}

\noindent where $l_t$ is the smoothed level of the series at instant $t$, $b_t$ denotes an estimate of the trend of the series at time $t$, $\alpha^*$ is a smoothing parameter for the level and $\beta^*$  is the smoothing parameter for the trend.
\smallskip

The application of Holt's method requires smoothing parameters $\alpha^*$, $\beta^*$, and initial values $l_0$ and $b$. All forecasts can be computed from the data after computing the initial values. Initial values are selected using the minimization of the sum of squared errors (SSE), defined as follows:

\begin{equation}
    \label{eq:sse}
    SSE=\sum_{t=1}^{T}(y_t-\hat{y}_{t|t-1})^2
\end{equation}

This formulation requires a nonlinear optimization. The implementation of the \texttt{forecasting} package in R (\citealp{Hyndman_24}) is adopted in this work.

\subsection{Forecasting with Neural Networks (NN)}

\textcolor{blue}{NN models can operate as universal approximators of nonlinear functions by learning complex relationships between the response variable and its predictors (\citealp{Scardapane2024}). NN models are made up of an input layer, hidden layers, and an output layer. The input for time-series forecasting is comprised of historical time-series signals up to a input window size $p$, $\bm{x}=\{x_t, x_{t-1},\ldots,x_{t-p+1}\}$. The hidden layer is defined as follows:}

\begin{equation}
   \textcolor{blue}{ h^{(l)}=\sigma(\bm{W}^{(l)}h^{(l-1)}+b^{(l)})}
\end{equation}

\noindent \textcolor{blue}{where $h^{(l)}$ is the output of the l-th layer (for the first layer $h^{(0)}=\bm{x}$), $\bm{W}^{(l)}$ is the weight matrix of the l-th layer, $b^{(l)}$ is a bias vector of the l-th layer, and $\sigma(.)$ is a nonlinear activation function.}

\textcolor{blue}{Finally, the output layer produces a $h$ steps ahead forecast, $\hat{x}_{t+h}$, chaining hidden layers and the output layer:}

\begin{equation}
    \textcolor{blue}{\hat{x}_{t+h}=\bm{W}^{(L)}h^{(L-1)}+b^{(L)}}
\end{equation}

\noindent \textcolor{blue}{where $L$ is the total number of layers.}

Forecasting with NNs requires the specification of a NN architecture that can approximate and extrapolate the underlying data-generating process. Namely, for time-series forecasting, it is necessary to evaluate the forecasting information of lagged signals and, if present, include seasonality. Accordingly, feature selection for time-series forecasting with NN needs to consider autoregressive and seasonality components.

In this context, to improve the accuracy of the prediction through a set of NN prediction models, (\citealp{Kourentzes_14}) developed NN ensemble operators for the prediction of time series, with excellent prediction results in different prediction competitions (\citealp{wang2023forecast}). This approach combines multiple base models in an ensemble model. Through the \texttt{nnfor} package in R, this work implements ensemble forecasting models based on \textcolor{blue}{Neural Network} (E-NN) and \textcolor{blue}{Extreme Learning Machines} (E-ELM) models (\citealp{NNFOR}). ELMs mitigate the optimisation of NN. Namely, instead of attempting to tune all weights in a NN, they are left to their random initial values, except for weights in the output layer.

In all the E-NN and E-ELM model configurations, a single hidden layer is used (\citealp{Kourentzes_14}). Regarding the number of neurons, it was evaluated by five-fold CV, using configurations from 2 up to 20 neurons. First-level differences are calculated over the input signal and lagged signals are evaluated in predefined windows and best lagged signals are selected. The series is also evaluated for seasonality and, if it is stochastic, seasonal differences are added to the input of the model (\citealp{Kourentzes_14}).  We use 20 ensemble members that are combined with the median operator. The activation function for E-NN and E-ELM is a sigmoid function. The output layer of the E-ELM is trained using Lasso regression. Backpropagation is used to train adjust the weights of each neuron by calculating the gradient of the loss function.

\subsection{Temporal Fusion Transformers (TFTs)}

\begin{figure*}[!htb]
	\centering
	\includegraphics[width=0.95\columnwidth]{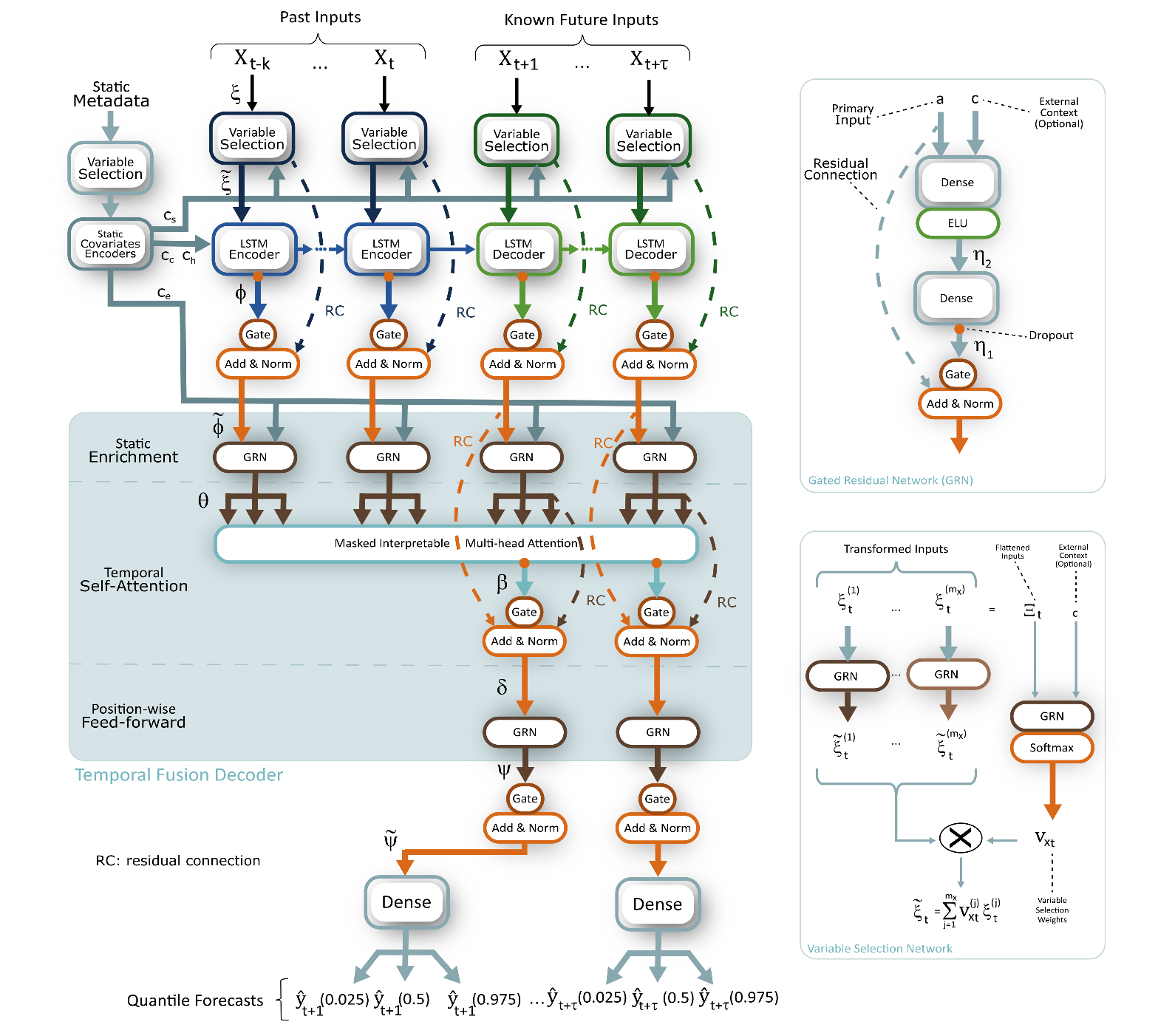}
	\caption{TFT architecture including Gated Residual Networks, Variable Selection Networks and encoder-decoder structure.}
	\label{fig:TFT_Architecture}
\end{figure*}

The architecture of TFTs was introduced in (\citealp{Lim_21}). TFTs are particularly effective for time series forecasting due to (i) their increased ability to capture long- and short-term dependencies through the self-attention mechanism; (ii) adaptive attention mechanism for dynamic temporal processing to prioritize important time points and ignore irrelevant information; (iii) ability to handle multiple data types including past observations, known future inputs and static covariates; and (iv) interpretable results which offer insights into which variables and time periods are most important for making predictions.

The TFT architecture is designed using canonical components to build feature representations for improved forecasting performance, which is shown in Figure~\ref{fig:TFT_Architecture}. This includes: 

\begin{enumerate}
\item[(a)]  \textit{gating mechanism}, which skips unused components, provides adaptive depth and network complexity; 
\item[(b)]  \textit{variable selection networks} to select relevant variables at each time step;
\item[(c)]  \textit{static covariate encoders} to integrate static features into the network through encoding context vectors to condition temporal dynamics;
\item[(d)]  \textit{temporal processing} to learn long and short term temporal relationships from the inputs. A sequence-to-sequence LSTM layer is employed for local processing and long-term dependencies are captured through a multi-head attention block; 
\item[(e)]  Finally, \textit{prediction intervals} via quantile  forecasts to determine the range of likely values for each prediction horizon.
\end{enumerate}

\textit{Gating mechanisms} are used to regulate the contribution of a sample, $a$, to a context vector $c$ through Gated Residual Networks (GRN), defined as follows:

\begin{equation}
    GRN_w(a,c)=\mathrm{Layer \ Norm(a+GLU_w}(\eta_1))
\end{equation}

\begin{equation}
    \eta_1=W_{1,w}\eta_2+b_{1,w}
\end{equation}

\begin{equation}
    \eta_2=ELU(W_{2,w}a+W_{3,w}c+b_{2,w})
\end{equation}

\noindent where ELU is the Exponential Linear Unit activation function, $\eta_1 \in R^{d_{model}}$ and $\eta_2 \in R^{d_{model}}$ are intermediate layers, LayerNorm is standard layer normalization, and $w$ is an index to denote weight sharing. The parameter $a$ represent the linear contribution. Component gating layers based on gated linear units (GLUs) are adopted to control the extent of nonlinear contribution to suppress unnecessary information in a given data set. The GLU is described as follows:

\begin{equation}
    GLU_w(\gamma)=\sigma(W_{4,w}\gamma+b_{4,w}) \odot (W_{5,w}\gamma+b_{5,w})
\end{equation}

\noindent where $\gamma$ is the input, $\sigma$ is the sigmoid activation function, $W_{(.)} \in R^{d_{model} \times d_{model}}$ and $b_{(.)} \in R^{d_{model}}$ are the weights and biases, $ \odot$ is the element-wise Hadamard product, $d_{model}$ is the hidden state size.

\textit{The variable selection networks (VSN)} ignore noisy inputs and concentrate on valuable data to improve model performance. VSN select the most relevant features from the input time-series data according to the forecasting horizon through a weighing mechanism. The variable selection weights provide interpretability of input variable for the TFT model. Let $\xi_t^{(j)}$  denote the transformed input of the j-th variable at time $t$, with being the flattened vector $\Xi_t=[\xi_t^{(1)^T}, \ldots, \xi_t^{(m_x)^T}]^T$ of all past inputs at time $t$. The flattened vectors and the static covariate $c_s$ are fed into a GRN and pass through a Softmax layer to calculate the selection weight of the variable.

\begin{equation}
\label{eq:1_VSN}
    v_{X_t}=\mathrm{Softmax}(GRN_{V_x}(\Xi_t,c_s))
\end{equation}

\begin{equation}
\label{eq:2_VSN}
    \xi_t^{j}=GRN_{\xi_t^{j}}(\xi_t^{(j)})
\end{equation}

\begin{equation}
\label{eq:weight_VSN}
    xi_{t}=\sum_{j=1}^{m_x}v_{x_t}^{(j)}\xi_t^{(j)}
\end{equation}

\noindent where $v_{X_t}^{(j)}$ is the j-th element of the vector $v_{x_t}$. The variable selection weight provides the explanatory properties for the forecasting of the TFT model, \textit{i.e.} the globally significant variables of the forecasting task can be identified in Eq.~(\ref{eq:2_VSN}), each is non-linearly processed by its GRN. Finally, the processed features are weighted by their variable selection weights and combined as shown in Eq.~(\ref{eq:weight_VSN}).

\textit{The attention function} can be summarized as mapping three vectors, a query and a key–value pair, to an output that is expressed as follows:

\begin{equation}
    \mathrm{Attention}(Q,V,K)=\mathrm{Softmax}(\frac{QK^T}{\sqrt{d_k}})V
\end{equation}

\noindent where $d_k$, is the dimension of $Q$, $Q \in R^{n\times d_k}$ is the query, $K \in R^{m\times d_k}$, and $V \in R^{m\times d_v}$ are the key-value data pair. \textcolor{blue}{The attention function} can be seen as a mapping sequence from $Q \in R^{n\times d_k}$ to  $Q \in R^{n\times d_v}$  by a dot product of  three matrices with sizes $n\times d_k$, $d_k \times m$ and $m\times d_v$.

Self-attention is a commonly used attention mechanism to indicate the attention distribution of a time-sequence. \textcolor{blue}{Self-attention} can be defined as $Attention(X,X,X)$, where $X \in R^{n\times d_k}$. Performing several attention functions in parallel with different projections is more beneficial than using a single attention function. Accordingly, multi-head attention is adopted to jointly yield the outputs from different mapping subspaces, which is expressed as:

\begin{equation}
    \mathrm{Multihead}(Q,V,K)=[H_1,\ldots,H_{m_H}]W_H
\end{equation}

\begin{equation}
    H_h=\mathrm{Attention}(QW_Q^{(h)},KW_K^{(h)},VW_V^{(h)})
\end{equation}

\noindent where $W_Q^{(h)}$, and $W_K^{(h)}$ $\in  R^{d_{model}\times d_{attn}}$, $W_V^{(h)} \in R^{d_{model}\times d_v}$, are head-specific weights for the keys, queries and values, and $W_H \in R^{d_{model}\times (d_vm_H)}$  linearly combines outputs concatenated from all the heads, $H_h$.

Given that each head uses a different value, the individual attention weight cannot represent the importance of a particular feature. The multi-head attention is thus changed to the shared value in each head:

\begin{equation}
    \mathrm{Interpretable \ Multi \ Head}(Q,K,V)=\accentset{\halfapprox}{H}W_H
\end{equation}

\begin{equation}
    \accentset{\halfapprox}{H}=\accentset{\halfapprox}{A}(Q,K)VW_V
\end{equation}

\begin{equation}
    \accentset{\halfapprox}{H}=\frac{1}{m_H}\sum_{h=1}^{m_H}\mathrm{Attention}(QW_Q^{(h)},KW_K^{(h)},VW_V^{(h)})
\end{equation}

The overall TFT architecture is framed with an LSTM encoder-decoder to capture the short-term memory, and the multi-head attention captures the long-term dependencies. The temporal fusion decoder, uses a series of (i) locality enhancement with sequence-to-sequence layers, (ii) static enrichment layer, (iii) temporal self-attention layer, and (iv) position-wise feedforward layer; to learn temporal relationships present in the dataset (\citealp{Lim_21}).

The encoder-decoder structure enables the static metadata to influence local processing. This is formalised through context vectors, $c_c,c_h$, static covariate encoder context $c_e$, and an external context vector $c_s$. Gated skip connection layers are applied and a static enrichment layer that enhances temporal features with static metadata.

After this stage, the self-attention is applied through static enriched temporal features and interpretable multi-head attention. Additionally, a gating layer is applied to facilitate training, which results in $\delta(t,n)$. Subsequently, a non-linear processing is applied to the outputs of the self-attention layer:

\begin{equation}
    \psi(t,n)=GRN_{\psi}(\delta(t,n))
\end{equation}

As shown in Figure \ref{fig:TFT_Architecture}, a gated residual connection is also applied, which results in $\accentset{\halfapprox}{\psi}$. Finally, quantile forecasts are generated using a linear transformation of the output of the temporal fusion decoder, $\accentset{\halfapprox}{\psi}(t,\tau)$, as follows:

\begin{equation}
    \hat{y}(q,t,\tau)=W_q\accentset{\halfapprox}{\psi}(t,\tau) +b_q
\end{equation}

\noindent where $W_q \in \mathbb{R}^{1\times d}$, $b_q\in \mathbb{R}$ are linear coefficients of the quantile $q$.

Finally, the TFT is trained by jointly minimizing the quantile loss function, summed across all quantile outputs:

\begin{equation}
\label{eq:Loss_TFT}
    \mathcal{L}=\sum_{y_t\in \Omega}\sum_{q\in Q}\sum_{\tau=1}^{\tau_{max}}\frac{QL(y_t,\hat{y}(q,t-\tau,\tau),q)}{M\tau_{max}}
\end{equation}

\begin{equation}
\label{eq:Loss_TFT2}
QL(y,\hat{y},q)=q(y-\hat{y})_m + (1-q)(\hat{y}-y)_m
\end{equation}

\noindent where $\Omega$ is the training domain, which includes $M$ samples, $W$ represents TFT weights, $Q$ is the set of quantiles ($Q=\{0.025,0.5,0.975\}$ in this work), and $(.)_m=max(0,.)$.

Transformers are known to perform very well on long-term time series (\citealp{Biggio_23}). In this research, the TFT package \texttt{Darts} available for Python is used (\citealp{Darts_22}). Overall, TFT shows very good generalization performance to unseen degradation conditions, and it is able to handle arbitrarily complex ageing profiles. This is the case even when these are chosen significantly outside the training distribution.  There are several hyperparameters that need to be considered when using the TFT. Consequently, different architectures were tested to select the best combination of three of those hyperparameters, including hidden size, LSTM layers and attention heads, as shown in Table~\ref{table:architecturesConfig}.

\begin{table}[htb]
    \centering
 \begin{threeparttable}
    \caption{\textcolor{blue}{Tested TFT hyperparameters.}}
    \setlength{\tabcolsep}{4.7pt}
        \begin{tabular}{m{6em} m{6em} m{6em} m{7em} }
             \toprule
             \textbf{Model name} & \textbf{Hidden size} & \textbf{LSTM layers} & \textbf{Attention heads} \\
             \midrule
             A & 128 & 4 & 4 \\ 
             B & 128 & 8 & 4 \\ 
             C & 128 & 4 & 2 \\ 
             D & 256 & 4 & 4 \\ 
             E & 256 & 8 & 4 \\ 
             F & 256 & 4 & 2 \\ 
            G & 64 & 1 & 4 \\
            \bottomrule           
        \end{tabular}
        \label{table:architecturesConfig}
 \end{threeparttable}
\end{table}	

In order to determine the best TFT architecture for different prediction horizons, the performance of LSTM layers, number of neurons, and attention heads were tuned.

\section{Case Study}
\label{sec:CaseStudy}

The MOSFET dataset provided by NASA (\citealp{Celaya_12}) is used to evaluate forecasting ability of different models in different configurations, time-scales, and learning proportions.

\subsection{\textcolor{blue}{Experimental Environment}}

\textcolor{blue}{Accelerated aging experiments are used to assess reliability in a much shorter time than long-term reliability tests. Specifically, thermal cycles are applied to create thermomechanical stresses in the switching devices due to the mismatch in the thermal expansion coefficients of different package elements. Failure conditions include latch-up, thermal runaway, or failure to turn on due to gate damage. In thermal cycles, no heatsink is used, significantly reducing heat dissipation, allowing for self-heating during commutes. The case temperature of the semiconductor device was measured using a thermocouple attached to the copper case flange (drain).}

\medskip

\textcolor{blue}{In this case, the accelerated thermal stress procedures were implemented for IRF520NPbF MOSFETs in run-to-failure experiments. For thermal cycling, on/off cycles were produced by applying a gate voltage of 15 V with a PWM signal at a frequency of 1 kHz and a duty cycle of 40\%. The drain-source was biased at 4 V DC, and a resistive load of 0.2 $\Omega$ was used on the source output. The dataset included gate-source voltage ($V_{GS}$), drain-source voltage ($V_{DS}$), drain current ($I_D$), and flange (case) temperature ($T_f$). The sampling rate of these time series is 1 MHz, with a transient response measurement every 400 ns.}

\medskip

\textcolor{blue}{X-ray and scanning acoustic images after degradation show that the die-attach solder migrates, forming voids. Die-attach damage due to thermal cycling reduces the contact area between solder-copper and solder-silicon interfaces, decreasing heat flow compared to a pristine die-attach. This results in a higher junction temperature and an increased ON resistance for the degraded device. Hence, the drain-to-source resistance ($R_{DS_{ON}}$) can be considered a precursor to assess the state of health (\citealp{Celaya_12}).  Furthermore, junction temperature is also a function of the case temperature, and the measured $R_{DS_{ON}}$ was normalized to eliminate the case temperature effects and reflect only changes due to degradation.}

\medskip

\textcolor{blue}{Due to manufacturing variability, not all transistors show an identical starting value of $R_{DS_{ON_0}}$ at time $t=0$, which is defined as pristine condition. For fair comparison, the initial $R_{DS_{ON_0}}$ value (“pristine condition value”) is subtracted from the absolute measurements. The normalized time series results in a trajectory ($\Delta R_{DS_{ON}}$) from pristine condition to failure, representing the degradation process due to die-attach failure and the increase in $R_{DS_{ON}}$ through the aging process.}
\medskip

\textcolor{blue}{Therefore, for prognostics, it is assumed that die-attach failure is the only active degradation mechanism during the accelerated aging experiment. The single value of $\Delta R_{DS_{ON}}$ is used to assess the health state of the device, with a failure threshold of a 0.05 increase in $\Delta R_{DS_{ON}}$ (\citealp{Celaya_12}).}

\subsection{Pre-processing Experiments}

\medskip

The main pre-procesing stages are divided into data sampling, temperature filtering, resistance normalization and temporal filtering. Namely, during the data sampling stage, firstly, \textcolor{blue}{the gate voltage } $V_{GS}$ is used to separate ON and OFF states of the MOSFET. Figure~\ref{fig:Cycling_Image_test36} shows the ON and OFF cycles of one experiment. \textcolor{blue}{The on-resistance $R_{DS_{ON}}$, which correlates with the junction temperature, is computed as the ratio of $V_{DS}$ to $I_D$ during the on-state (with duration $T_{ON}$) of the square waveform (\citealp{Celaya_12}):}
\medskip

\begin{equation}
\label{eq:RDS_On}
    R_{DS_{ON}}(k)=\sum_{i=1}^{T_{ON_i}}\frac{V_{DS}(i)}{I_D(i)}
\end{equation}

\noindent where $k=\{T_{ON_1},\ldots,T_{ON_K}\}$ are the different ON cycles during the MOSFET accelerated thermal ageing process, and $i=\{1,\ldots,T_{ON_i}\}$ is the duration of i-th ON cycle of the MOSFET. 
\smallskip

Subsequently, \textcolor{blue}{$R_{DS_{ON}}$ values computed at flange temperature values below the \textit{$T_f<$Low Temp} threshold were removed from the data in order to discard $R_{DS_{ON}}$ values with no influence on thermal ageing. Furthermore, in order to eliminate the case temperature effects and observe only the changes produced by degradation, resistance normalization was carried out subtracting the pristine condition to all the $R_{DS_{ON}}$ values (\citealp{Celaya_12}).} Finally, minute-based mean values of $\Delta R_{DS_{ON}}$ values were computed to filter out noise.

\begin{figure}[!h]
	\centering
	\includegraphics[width=0.5\columnwidth]{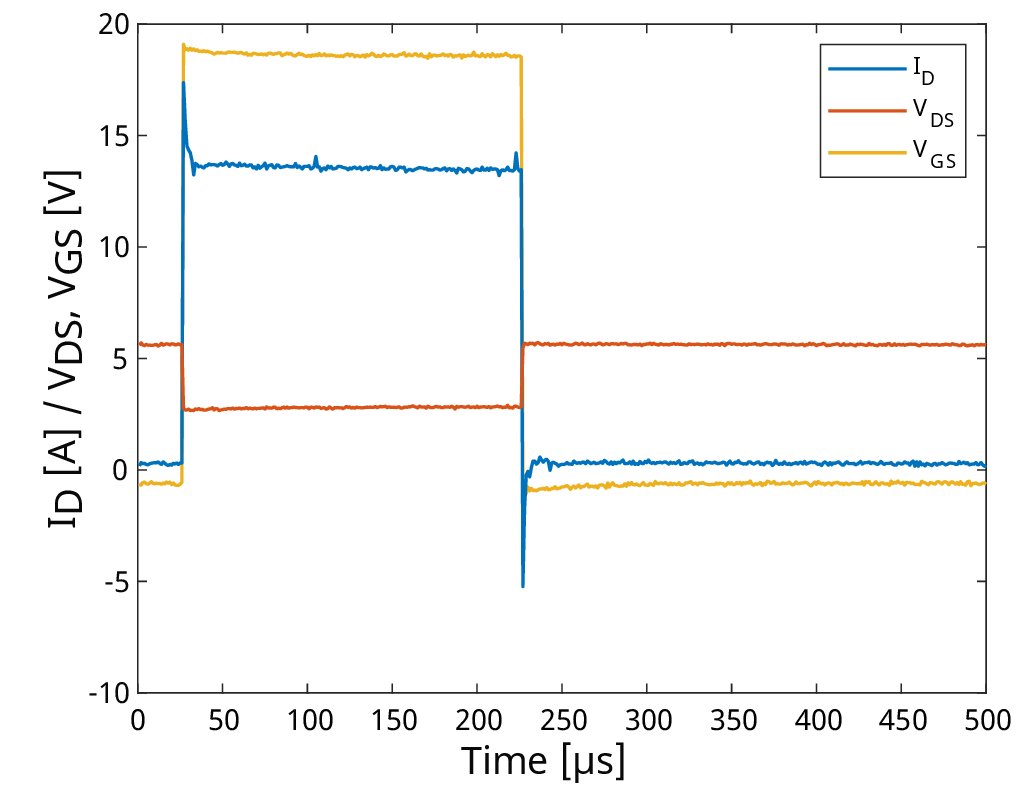}
	\caption{$V_{DS}$, $V_{GS}$ and $I_{D}$ used to compute $R_{DS_{ON}}$.}
	\label{fig:ResultingTests}
\end{figure}

Figure~\ref{fig:ResultingTests} shows the calculation example of $R_{DS_{ON}}$, \textcolor{blue}{including $V_{DS}$, $V_{GS}$ and $I_{D}$, that are used to compute $R_{DS_{ON}}$ [cf. Eq.~(\ref{eq:RDS_On})]}.
\smallskip

\begin{figure}[!h]
	\centering
	\includegraphics[width=0.55\columnwidth]{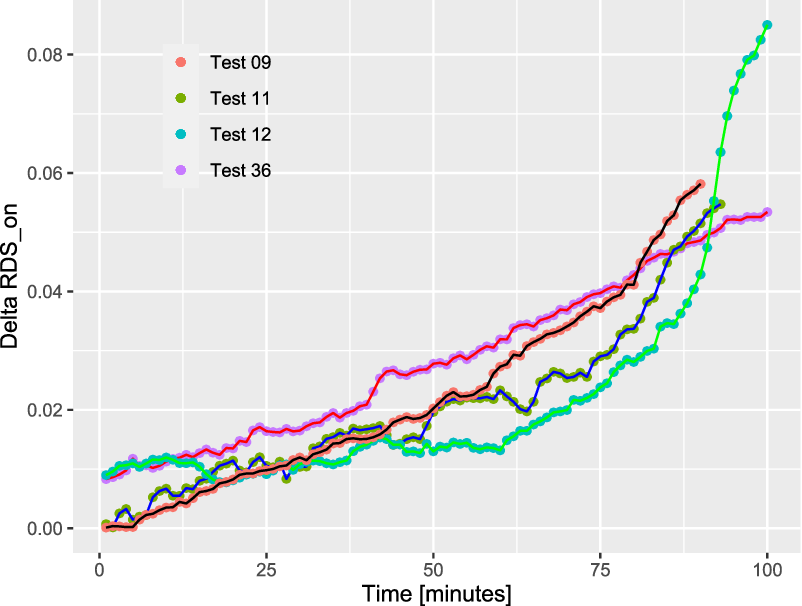}
	\caption{Obtained run-to-failure experiments after pre-processing.}
	\label{fig:ResultingTests_RunFail}
\end{figure}

\textcolor{blue}{After this process, four experiments were selected from the total of 42 experiments. This selection process was done based on the amount of remaining data points for each experiment after completing the pre-processing stage and evaluating the instant when failure condition was reached. Accordingly, Figure~\ref{fig:ResultingTests_RunFail} shows the available run-to-failure experiments of the MOSFETs after the pre-processing stage.}

\textcolor{blue}{Figure~\ref{fig:ResultingTests_RunFail} shows that there is an inherent variability in the ageing processes of the same MOSFETs for different tests. This variability may have implications when designing an ageing forecasting model for MOSFETs of the same family.}

\subsection{Statistical properties \& Evaluation Metrics}

All the available experiments are non-stationary, which have been evaluated through the Dickey-Fuller test. In addition, through the analysis of the autocorrelation, it has been observed that all the experiments have a trend and they do not have seasonality. These characteristics make the experimental datasets good candidates to apply ARIMA and Holt models. 
\smallskip

The evaluation metric Mean Average Percentage Error (MAPE) is defined as follows:

\begin{equation}
    \label{eq:mape}
    MAPE=\frac{100}{N}\sum_{t=1}^{t=N}|\frac{y_t-\hat{y_t}}{y_t}|
\end{equation}

\section{Numerical Experiments}
\label{sec:NumericalExperiments}

In this section, MOSFET ageing forecasting models have been designed and tested for short- and long-term predictions.

\subsection{Short-term MOSFET Ageing Forecasts}
\label{ss:Prognostics_Filters}

\subsubsection{Model design}

The performance of the selected model architectures was evaluated by training and testing in an iterative way. Given a prediction horizon of $N$ steps ahead, the process starts training with the first $p$ samples of a given experiment and makes a $1$ step-ahead prediction. If $N>1$,  then, forecast results are concatenated with the training set and thus, the training set is incremented in $N$. Then again, another $N$ step ahead prediction is performed. This process is repeated until the prediction of the last available point of the experiment is carried out. Algorithm \ref{alg:ShortTerm} describes the details of the training-testing process. Initial tests have been performed using 33 \% of the available dataset for each experiment.

\begin{algorithm}[!htb]
	\caption{Short-term training.}
 \small
	\label{alg:ShortTerm}
\begin{algorithmic}[1]
	\Statex \textbf{Inputs:}
        \Statex ${D} = [d_1, \ldots, d_K]$ \Comment{Data vector of $K$ elements.} 
        \Statex $p$  \Comment{Data-points to fit the model, $p<K$.} 
	\Statex $g(.)$  \Comment{Forecasting model.}
        \Statex $D_{T} = [d_1, \ldots, d_{p}]$  \Comment{Training vector, $p$ elements.}
	\Statex \textbf{Output:} ${\mathcal{V}}$ \Comment{Predictions vector.} 
	\Statex $h=K-p$ \Comment{Prediction horizon.}
	\For{$(i=1:h)$} \Comment{For data length.}
        \State $g=fit(D_{T})$ \Comment{Fit model $g$ with $p$ data-points.}
        \State $\mathcal{F}=g(D_{T},1)$ \Comment{$1$-step ahead forecast.}
        \State $\mathcal{V}  \leftarrow \mathcal{F}$ \Comment{Save forecast in $V$}
        \If{$N > 1$} \Comment{Multi-step ahead forecast.}
            \For{$(j=2:N)$}
                \State $D_{T} = D_{T} \cup \mathcal{F}$ \Comment{Concatenate.}
                \State $g=fit(D_{T})$ \Comment{Re-fit model $g$.}
                \State $\mathcal{F}=g(D_{T},1)$ \Comment{$1$-step ahead forecast.}
                \State $\mathcal{V}  \leftarrow \mathcal{F}$ \Comment{Save forecast in $V$}
            \EndFor     
        \EndIf
        \State $p = p + N$ \Comment{Increase $p$ in $N$}
        \State $D_{T} = [d_1:d_{p}]$ \Comment{Update training vector.} 
        \EndFor
        \State \textbf{return}$(\mathcal{V})$
\end{algorithmic}
\end{algorithm}
\normalsize

\begin{figure}[!h]
    \centering
    \includegraphics[width=0.6\columnwidth]{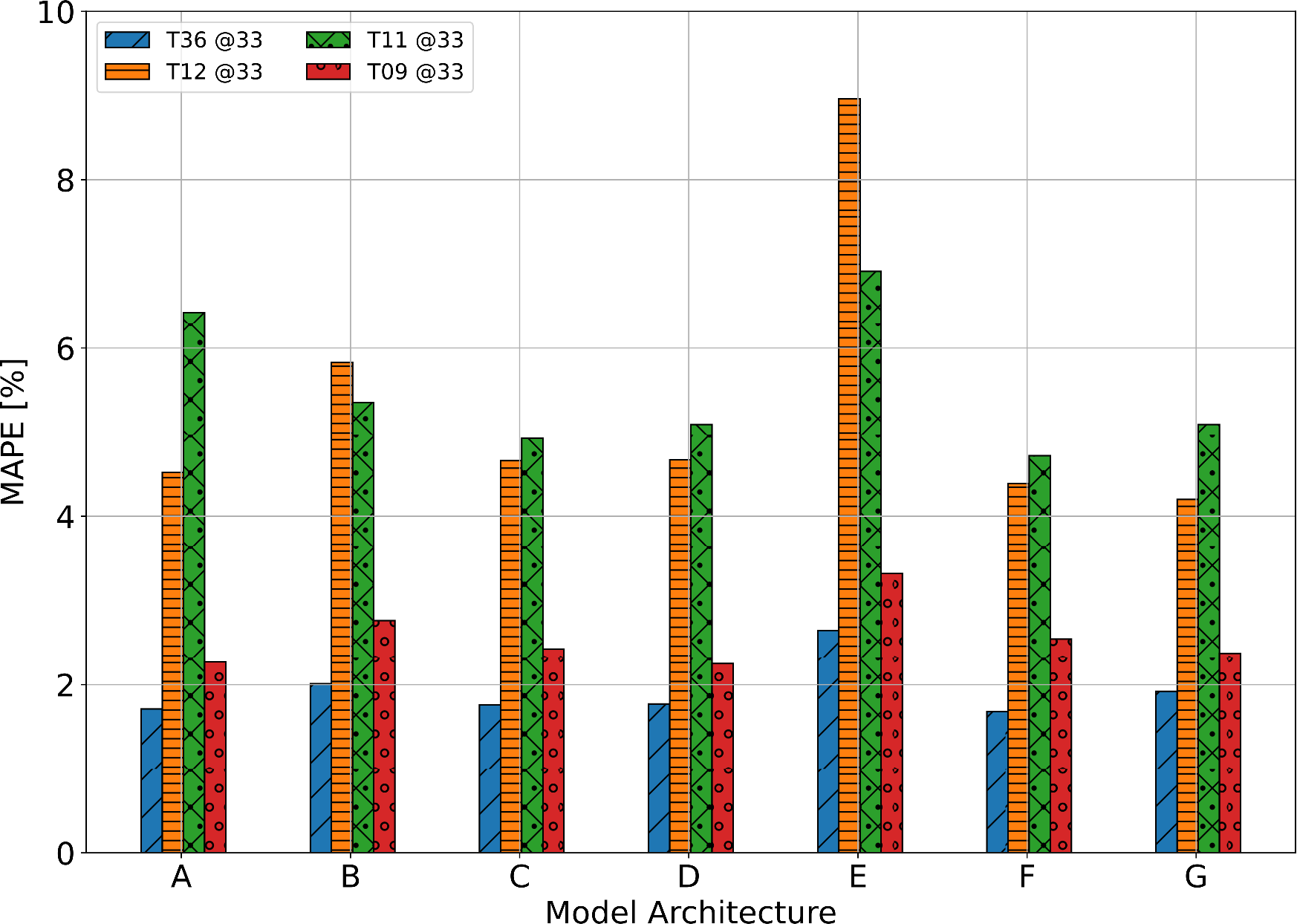}
    \caption{TFT architecture selection for short-term MOSFET ageing forecasting for different tests \textcolor{blue}{(T09, T11, T12, T36)} with 33\% of data used for training.}
    \label{fig:shortterm_ArchSel}
\end{figure}

For each of the forecasting models (EKF, UKF, ARIMA, Holt, TFT, E-ELM, E-NN), in \texttt{Line 2} of the Algorithm \ref{alg:ShortTerm} it is necessary to adjust the hyperparameters to fit the model. In the TFT model, a hyperparameter tuning process is needed to select to adjust the parameters (see Table~\ref{table:architecturesConfig}). The obtained results are presented in Figure~\ref{fig:shortterm_ArchSel}.

\textcolor{blue}{Figure~\ref{fig:shortterm_ArchSel} shows that there is a direct dependence between the hyperparameters and obtained results.} The best TFT architecture was selected by counting the number of times that each architecture performed best among all solutions. \textcolor{blue}{In this case, the model G performed best among all configurations (cf. Table~\ref{table:architecturesConfig})}. The hyperparameter tuning process is repeated for all models with their respective parameters. After fitting the model, predictions are made iteratively across the validation set. Once the predictions are obtained, $\mathcal{V}$, the MAPE of the prediction results is computed. 

\subsubsection{Prediction Results}

Table~\ref{table:Prognostics_Results} displays the obtained mean MAPE results for MOSFET ageing predictions at different forecast horizons $N=\{1, 2, 4\}$ for different experiments and different predictive models using the available 33\% of the datasets as training sets.

\begin{table}[hbtp]
	\centering
	\caption{\textcolor{blue}{Mean $\Delta$R\textsubscript{DS\textsubscript{ON}} prediction MAPE for different tests, models and steps-ahead (N) --- best results highlighted.}}
	\label{table:Prognostics_Results}
	\begin{tabular}{
  >{\centering\arraybackslash}m{.6cm}|
>{\centering\arraybackslash}m{2cm}|
>{\centering\arraybackslash}m{1cm}|
>{\centering\arraybackslash}m{1cm}|
>{\centering\arraybackslash}m{1cm}} \hline
\multirow{2}{*}{\small{\textbf{Test}}} & \multirow{2}{*}{\small{\textbf{Model}}} & \multicolumn{3}{c}{\textbf{Forecasting Horizon}} \\ \cline{3-5}
 &  & \small{\textbf{N=1}} & \small{\textbf{N=2}}  & \small{\textbf{N=4}}  \\ \cline{1-5} \hline
   \multirow{6}{*}{\small{\textbf{\#36}}} &  \small{\textbf{EKF}} & \small{2.61} & \small{3.65}  & \small{5.44}  \\ 
  & \small{\textbf{UKF}} & \small{2.09} & \small{2.82}  & \small{4.58}  \\ 
        & \small{\textbf{HOLT}} & \small{1.65} & \small{1.84}  & \small{1.93}  \\ 
      & \small{\textbf{ARIMA}} & \small{1.6} & \small{1.93}  & \small{2.17}  \\ 
      & \small{\textbf{TFT}} & \small{1.92} & \small{2.51}  & \small{3.39}  \\ 
   & \small{\textbf{E-NN}} & \small{1.9} & \small{2.2}  & \small{3.1}  \\ 
   & \small{\textbf{E-ELM}} &  \textbf{\small{1.6}} & \textbf{\small{1.8}}  & \textbf{\small{1.9}}  \\ \cline{1-5} \hline
\multirow{6}{*}{\small{\textbf{\#9}}} & \small{\textbf{EKF}} & \small{3.39} & \small{4.47}  & \small{7.37} \\ 
  & \small{\textbf{UKF}} & 
  \small{3.31} & \small{4.67}  & \small{7.45}  \\ 
         & \small{\textbf{HOLT}} & \small{2.19} & \small{2.55}  & \textbf{\small{3.47}}  \\ 
         & \small{\textbf{ARIMA}} & \small{2.04} & \small{2.5}  & \small{3.56}  \\ 
         & \small{\textbf{TFT}} & \small{2.37} & \small{3.35}  & \small{5.33}  \\ 
   & \small{\textbf{E-NN}} & \small{2.3} & \small{3}  & \small{5}  \\ 
   & \small{\textbf{E-ELM}} & \textbf{\small{2}} & \textbf{\small{2.5}}  &  \small{3.9}  \\ \cline{1-5} \hline
\multirow{6}{*}{\small{\textbf{\#11}}} & \small{\textbf{EKF}} & \small{7.12} & \small{11.1}  & \small{14.3} \\
  & \small{\textbf{UKF}} &  \small{4.38} & \small{6.39}  &   \small{8.44} \\ 
         & \small{\textbf{HOLT}} & \small{4.51} & \small{5.60}  & \small{7.80}  \\ 
         & \small{\textbf{ARIMA}} & \textbf{\small{3.86}} &  \textbf{\small{5.48}}  &  \textbf{\small{7.57} } \\ 
         & \small{\textbf{TFT}} & \small{5.09} & \small{6.73 }  & \small{10.5}  \\ 
   & \small{\textbf{E-NN}} & \small{5.1} & \small{6.8}  & \small{9.8} \\ 
   & \small{\textbf{E-ELM}} & \small{4.5} &   \small{6.1}  & \small{8.8} \\ \cline{1-5} \hline
   \multirow{6}{*}{\small{\textbf{\#12}}} &  \small{\textbf{EKF}} & \small{8.90} & \small{13.7}  & \small{18.2} \\ 
  & \small{\textbf{UKF}} & \small{5.54} & \small{6.38}  & \small{8.89} \\ 
& \small{\textbf{HOLT}} & \small{4.57} & \small{5.11}  & \small{8.17}  \\ 
& \small{\textbf{ARIMA}} &  \textbf{\small{3.68}} & \textbf{\small{5.02}}  &  \textbf{\small{7.15}}  \\ 
& \small{\textbf{TFT}} & \small{4.20} & \small{7.03}  & \small{11.1}  \\ 
   & \small{\textbf{E-NN}} & \small{5} & \small{6.8}  & \small{9.8}  \\ 
   & \small{\textbf{E-ELM}} &  \small{4.5} &  \small{6.1}  & \small{8.8}  \\ \cline{1-5} \hline
\end{tabular}	
\end{table}

Table~\ref{table:Prognostics_Results} shows that, consistently across different forecasting horizons, E-ELM performs better for tests \#9 and \#36, except N=4 for test \#9 with HOLT's model, which in practice is also a filter based on weighted past values, and for the tests \#11 and \#12, ARIMA performs best.

The E-ELM handles well tests \#9 and \#36 because it lacks of underlying model-form and noise hypotheses. It models the future evolution from a purely black-box perspective, including neuron weights and biases optimized to mimic the $R_{{DS}_{ON}}$ evolution, based on the selected delayed signals. In contrast, it can be observed that the tests \#11 and \#12 include more variability. Accordingly, this reflects the capability of the ARIMA model to deal with variable signals, filtering out local fast variations that have no relationship with long-term behaviour.

\begin{figure}[!h]
	\centering
	\includegraphics[width=\textwidth]{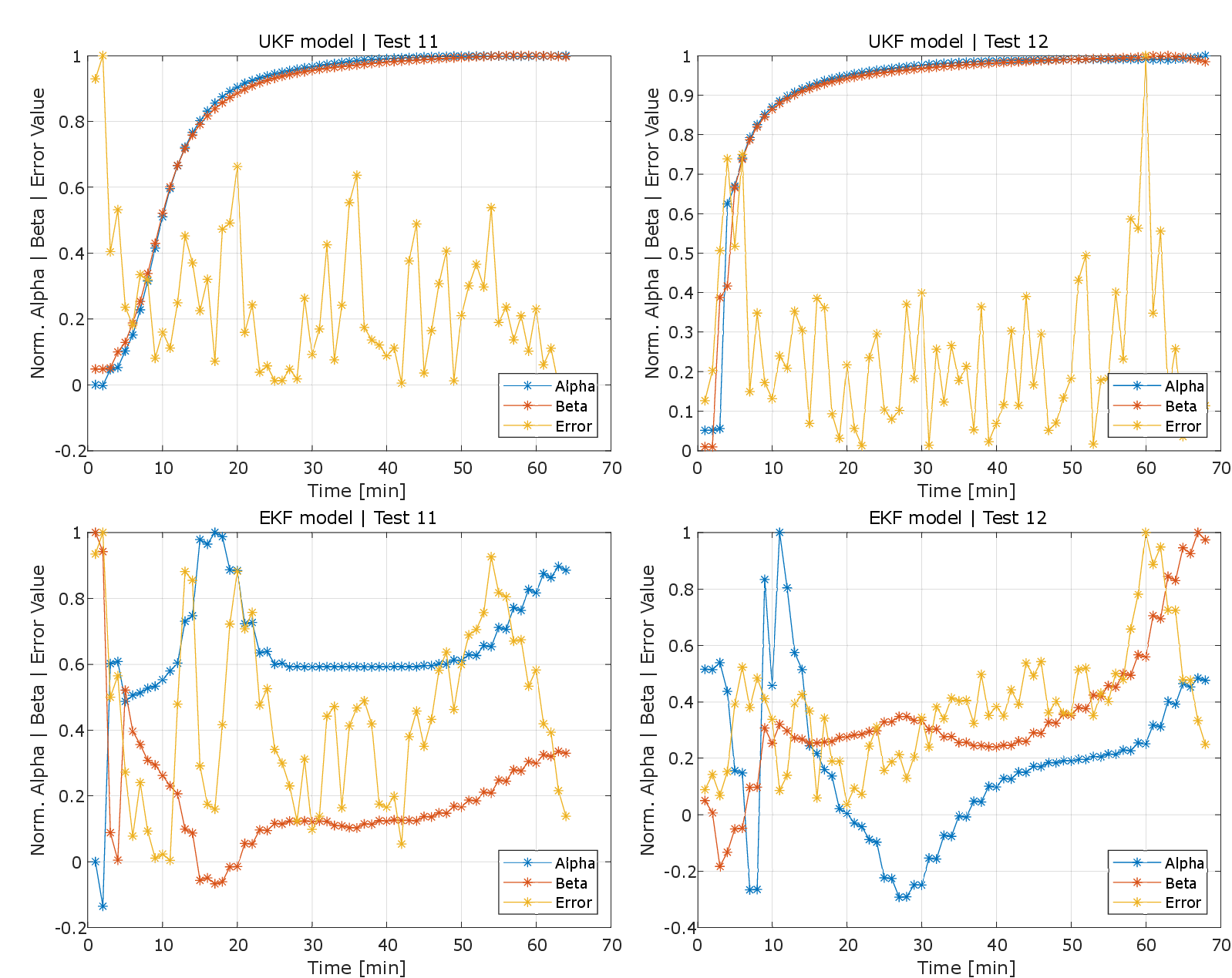}
	\caption{Adaptation of $\alpha$ and $\beta$ parameters with respect to prediction error for N=2 for the tests \#11 and \#12 for the UKF and EKF models.}
	\label{fig:AlphaBeta_N_2}
\end{figure}

Among state-space models, UKF is the most accurate for all tests and different prediction horizons. Although it does not obtain the most accurate results for the tests \#9 and \#36, UKF results are comparable to the rest of the models. In contrast, for the EKF model with tests \#11 and \#12 the accuracy drops substantially. Clearly in this case, the Taylor series approximation of the EKF was not sufficient to capture the effects the non-linearities present in the model. The EKF lacks the ability to adapt to changes in signal modifications, resulting in continuous oscillations of the state-space parameters $\alpha$ and $\beta$, which have a direct impact on the error. In contrast, UKF results are better because the parameters are more stable and adapt to signal modifications. Figure~\ref{fig:AlphaBeta_N_2} shows the influence of the modification of $\alpha$ and $\beta$ \textcolor{blue}{parameters in the EKF and UKF models} with respect to the learning horizon.

\textcolor{blue}{Figure~\ref{fig:AlphaBeta_N_2} shows that the UKF model parameters adapt and stabilize as the prediction time progresses, whereas the EKF model parameters are non-stable, aligned with the limitations of the EKF to match the nonlinear characteristics outside the operation point, and this is reflected in the increased prediction error.}

As for the TFT model for short-term predictions, it can be observed that the performance is in between the best and worst performing models. Increasing the forecasting horizon increases the MAPE of all models across different tests.

\begin{figure*}[!h]
	\centerline{
	\includegraphics[width=1.3\columnwidth]{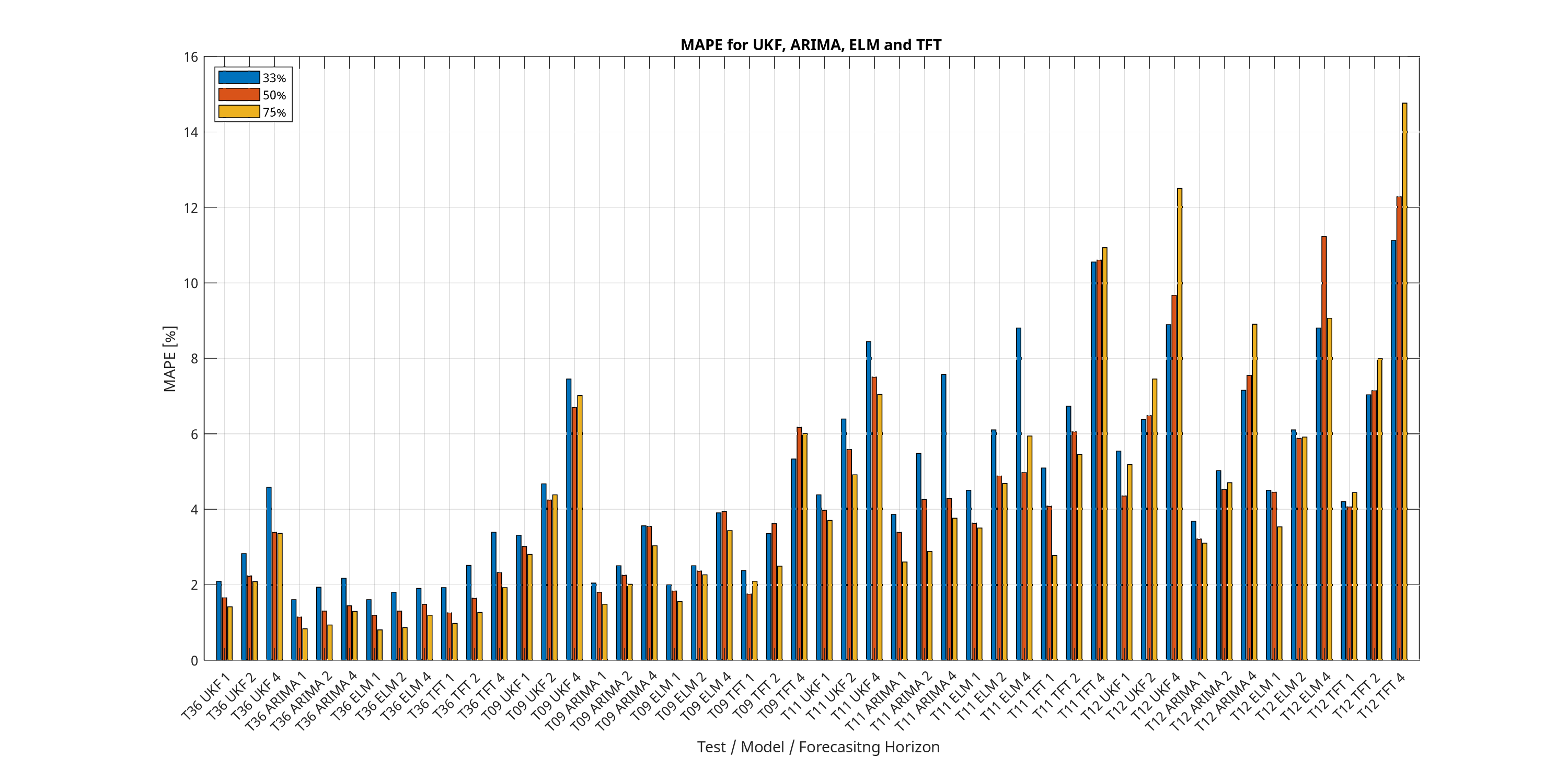}
 }
	\caption{MAPE values for different training proportions, tests and models.}
	\label{fig:RunToFailure_Tests}
\end{figure*}

\subsubsection{Influence of train-test proportions}

In order to evaluate the influence of different training-testing proportions on the forecasting accuracy, along with the TFT model, the best performing state-space statistical forecasting and NN with AR inputs models have been tested. They correspond to \textit{i.e.} UKF, ARIMA and E-ELM models for different tests and different forecasting horizons. Figure~\ref{fig:RunToFailure_Tests} shows the obtained results.

Figure~\ref{fig:RunToFailure_Tests} shows that, consistently across all models, the more training data, the lower the MAPE. Moreover, the longer the prediction horizon, the greater the MAPE. There are some exceptions, especially for the UKF model, where the training set determines the initial conditions. Namely, the corresponding $\alpha$ and $\beta$ values are calculated through least squares using the training dataset, and then these parameters are used as initial values for the state-estimation algorithms. This is not always an information gain for the state-space model based forecasting, as observed in test \#12.
\smallskip

As for the ARIMA and E-ELM, parameters of the trained models fully depend on the available data. Therefore, if longer datasets include the behaviour of the testing set, then the accuracy and uncertainty should be lowered. However, if the testing set includes out-of-distribution samples, the ability to model the ageing will become smaller.

\subsubsection{Influence of the state-transition model}

It is possible to test the state-space models with alternative definitions. Namely, Table~\ref{table:Progonstics_Results_EquationC} shows the mean MAPE results for the state transition model in Eq.~(\ref{eq:ss1_der2}) using 33\% of the datasets for training. 

\begin{table}[hbtp]
	\centering
	\caption{\textcolor{blue}{Mean $\Delta$R\textsubscript{DS\textsubscript{ON}} prediction MAPE for the UKF model.}}
	\label{table:Progonstics_Results_EquationC}
	\begin{tabular}{
  >{\centering\arraybackslash}m{.8cm}|
  >{\centering\arraybackslash}m{2cm}
>{\centering\arraybackslash}m{2cm}
>{\centering\arraybackslash}m{2.4cm}} \hline
\multirow{2}{*}{\small{\textbf{Test}}} & \multicolumn{3}{c}{\textbf{Steps-ahead}}   \\ \cline{2-4}
 &  \small{\textbf{N=1}} & \small{\textbf{N=2}}  & \small{\textbf{N=4}}  \\ \hline
\small{\textbf{\#9}} &  \small{3.49} & \small{5.02}  & \small{8.12}  \\ 
\small{\textbf{\#11}} &  \small{4.76} & \small{7.18}  & \small{10.00}  \\ 
\small{\textbf{\#12}}  &  \small{4.97} & \small{6.70}  & \small{9.99} \\ 
\small{\textbf{\#36}} &  \small{2.80} & \small{4.15}  & \small{7.19}  \\ \hline
\end{tabular}	
\end{table}

Comparing the UKF model results displayed in Table~\ref{table:Prognostics_Results} with Table~\ref{table:Progonstics_Results_EquationC}, which uses the state-space transition model in Eq.~(\ref{eq:ss2_der2}), it can be observed that the MAPE in the latter case is greater for all the different tests (except, test \#12, N=1), demonstrating the superior capabilities of the state-space model in Eq.~(\ref{eq:ss2_der2}) for modelling non-linear processes.

\subsection{Long-term MOSFET Ageing Forecasting}


In order to evaluate the generalisation capability of the designed models on unseen experiments, long-term forecasting models have been designed and tested. To this end, the TFT model architecture has been used, which has shown good results in various long-term prediction applications (\citealp{Biggio_23}). In addition, the best model from Table~\ref{table:Prognostics_Results} has been selected for comparison purposes, \textit{i.e.} the ELM model. 

Based on the assumption that a Leave One Out (LOO) testing is designed for MOSFETs of the same family, firstly, outlier experiments are discarded, which indicates different MOSFET technology. To this end, the Fréchet distance was calculated among different experiments (\citealp{har2014frechet}). Table~\ref{table:FrechetDistance} displays the obtained results.

\begin{table}[htb]
    \centering
    \caption{\textcolor{blue}{Similarity of experiments based on Fréchet distance.}}
    \setlength{\tabcolsep}{4.7pt}
        \begin{tabular}{m{4em} m{4em} m{4em} m{4em} m{4em} }
             \toprule
             \textbf{Test} & \textbf{\#9} & \textbf{\#11} & \textbf{\#12} & \textbf{\#36} \\
             \midrule
             \textbf{\#9} & 0 & 0.003 & 0.027 & 0.008 \\ 
             \textbf{\#11} & 0.003 & 0 & 0.03 & 0.008 \\ 
             \textbf{\#12} & 0.027 & 0.03 & 0 & 0.032 \\ 
             \textbf{\#36} & 0.008 & 0.008 & 0.032 & 0 \\ 
            \bottomrule           
        \end{tabular}
        \label{table:FrechetDistance}
\end{table}	

It is observed that the distance of test \#12 with respect to other tests is greater. This potentially indicates that it pertains to a different MOSFET family, and therefore, it is not considered for LOO experiments.

\subsubsection{Model design}


One of the strengths of the TFT is that it can incorporate covariates that impact on model predictions. Accordingly, to inform the long-term performance of the MOSFET aging, a covariate has been designed based on the piecewise linearisation of the training tests. 

The piecewise linearisation consists of (i) splitting the tests into blocks of 20\% data samples (ii) curve-fitting through least squares for each block and (iii) finally, the covariate is defined as a piecewise function of the fitted curves. This is described in Eq.~(\ref{eq:covariate}), where $\{m_1, \ldots, m_5\}$ and $ \{b_1, \ldots, b_5\}$, correspond to the linear curve parameters for the fitted curve.

\begin{equation}
    \label{eq:covariate}
    y(t) = \begin{cases}
        m_1 t + b_1 \: 0 \leq t \leq 19 \\
        m_2 t + b_2 \: 20 \leq t \leq 39 \\
        m_3 t + b_3 \: 40 \leq t \leq 59 \\
        m_4 t + b_4 \: 60 \leq t \leq 79 \\
        m_5 t + b_5 \: 80 \leq t \leq 99 \\
        \end{cases}
\end{equation}

\begin{figure}[h]
	\centering
	\includegraphics[width=0.5\columnwidth]{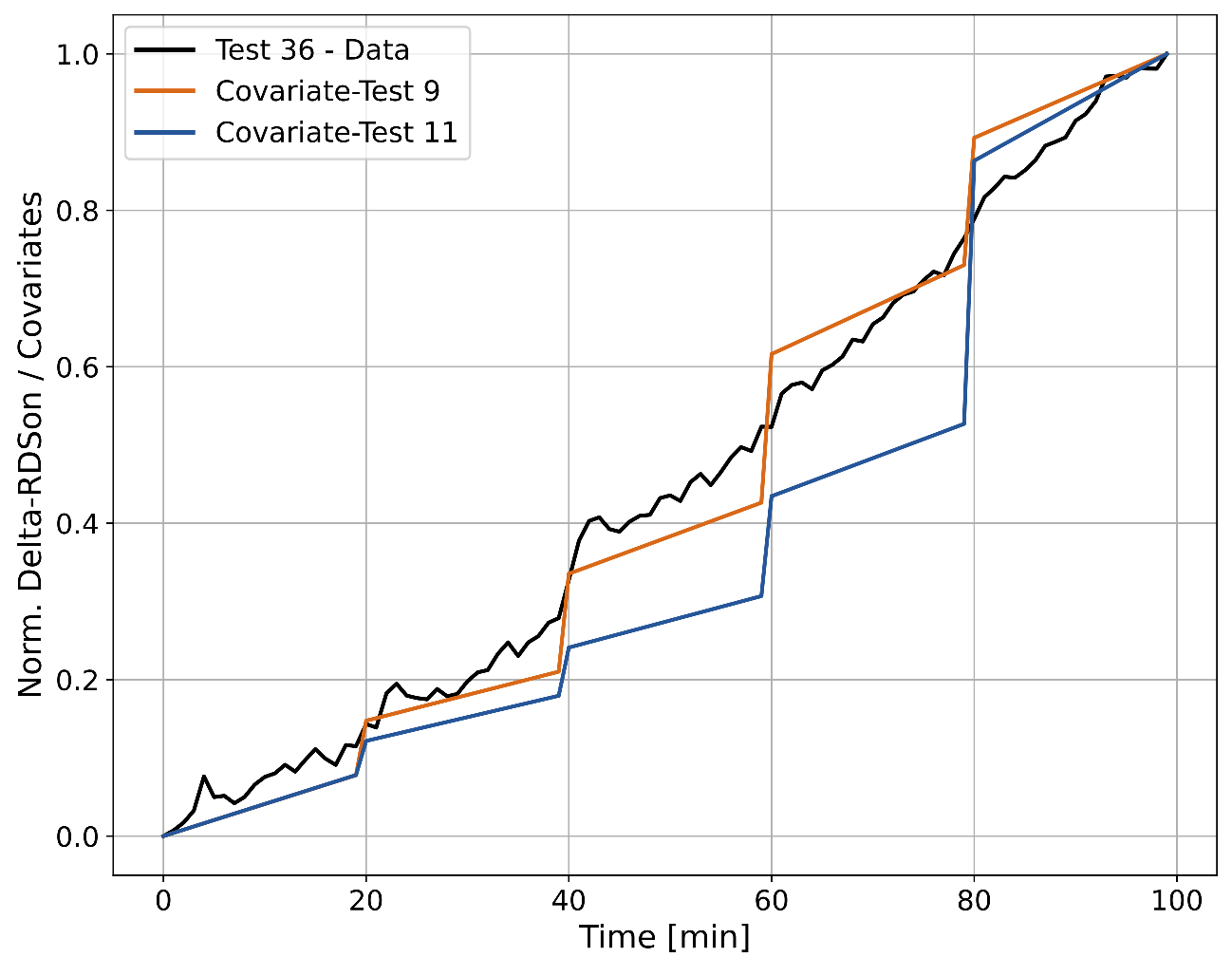}
	\caption{Normalized test 36 with the designed covariates using tests 9 and 11.}
	\label{fig:cov_T36}
\end{figure}

Figure~\ref{fig:cov_T36} shows an example of the covariates used for test 36 through the piecewise linearization of tests 9 and 11.



\begin{figure*}[!htb]
\centering
	\includegraphics[width=1\columnwidth]{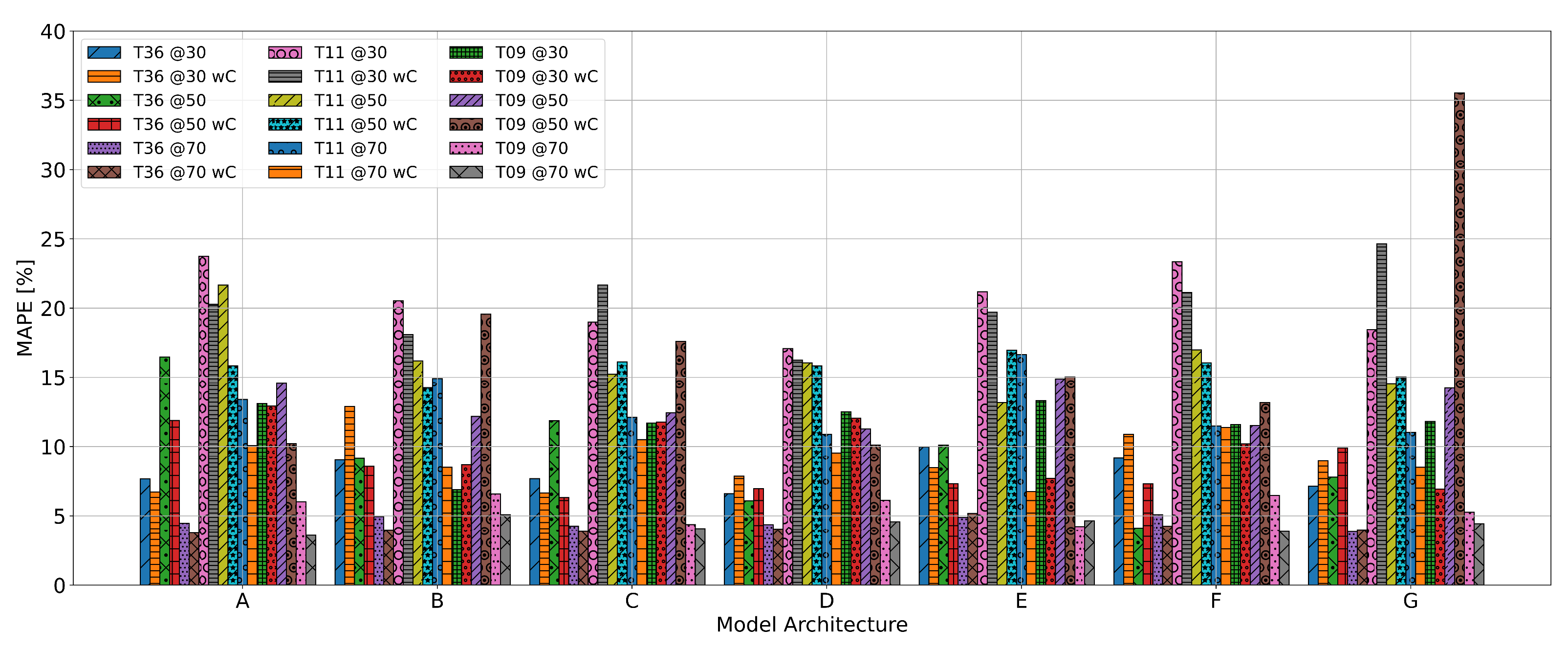}
	\caption{TFT architecture selection for long-term MOSFET ageing forecasting.}
	\label{fig:longterm_ArchSel}
\end{figure*}

To select the long-term architecture, different combinations of hyperparameters have been tested (cf. Table~\ref{table:architecturesConfig}). Figure~\ref{fig:longterm_ArchSel} shows the results of the selection of the model architecture \textcolor{blue}{for different tests (T09, T11, T12, T36), with different training data proportions (30, 50, 70), and including TFT models without and with covariates (wC).}

\textcolor{blue}{Figure~\ref{fig:longterm_ArchSel} again confirms that the TFT hyperparameters impact on the obtained results.} Among the architectures tested, the best TFT architecture was selected based on the best overall performance criteria. That is, it is selected by counting the number of times that each architecture performs best. It is observed that the best architecture is represented by model D.
\smallskip

\subsubsection{Prediction Results}

\begin{figure}[!h]
    \centering
    \includegraphics[width=0.6\columnwidth]{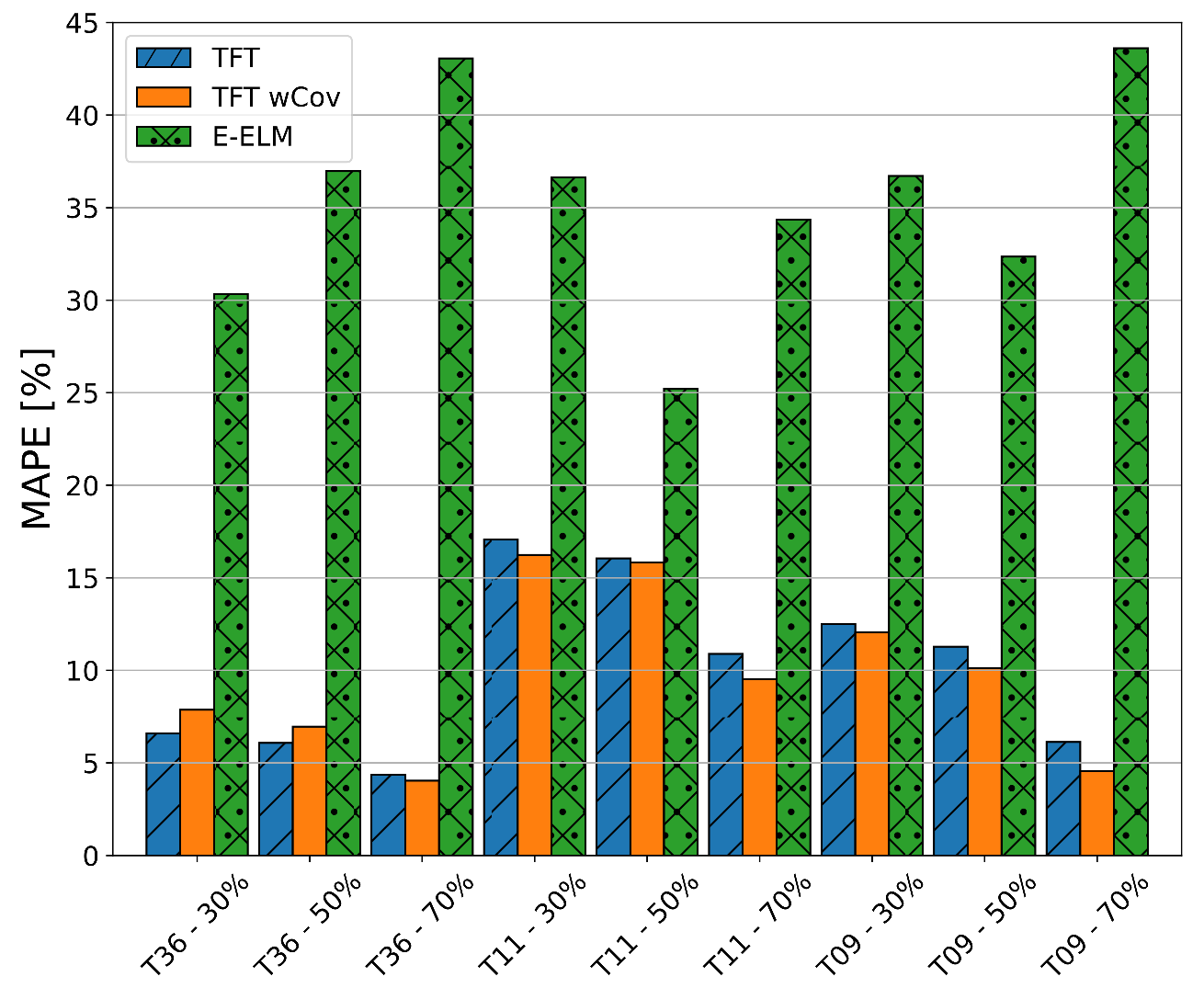}
    \caption{Long term $\Delta$R\textsubscript{DS\textsubscript{ON}} prediction MAPE for different tests, models and training proportions.}
    \label{fig:tftvselmlong}
\end{figure}

Finally, a LOO test was performed by training the model with 2 tests and leaving the third for the evaluation of the model. Accordingly, Figure~\ref{fig:tftvselmlong} shows the obtained results for different (a) models: TFT, TFT with covariate \textcolor{blue}{(TFT wCov)}, and ELM; (b) LOO evaluation experiments: T36, T11, and T09; and (c) training proportions: 30\%, 50\%, 70\% training samples, or equivalently, 70, 50 and 30 steps-ahead forecasting horizons, respectively.

\newpage

Figure~\ref{fig:tftvselmlong} \textcolor{blue}{shows} that the TFT obtains a much smaller MAPE than the ELM for all the different configurations. Among the TFT variants, the TFT with covariate is the model with the lowest MAPE, except in two configurations (T36, 30\% and 50\%). As for the MAPE with respect to the prediction horizon, for the TFT architectures, the MAPE is reduced for smaller forecasting horizons.

\begin{figure}[!htb]
	\centering
	\includegraphics[width=0.6\textwidth]{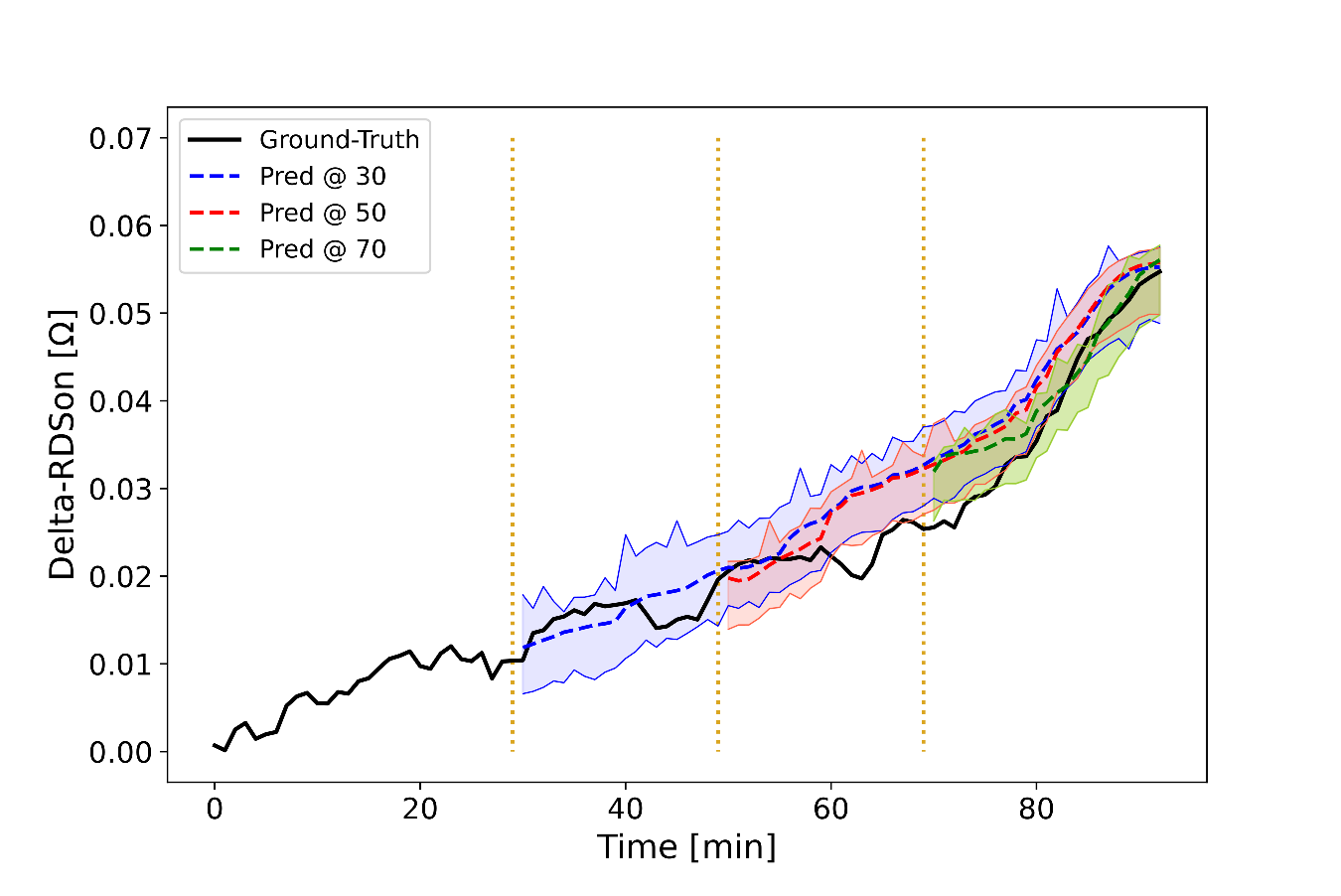}\\
 \vspace{-0.5cm}
        \includegraphics[width=0.6\textwidth]{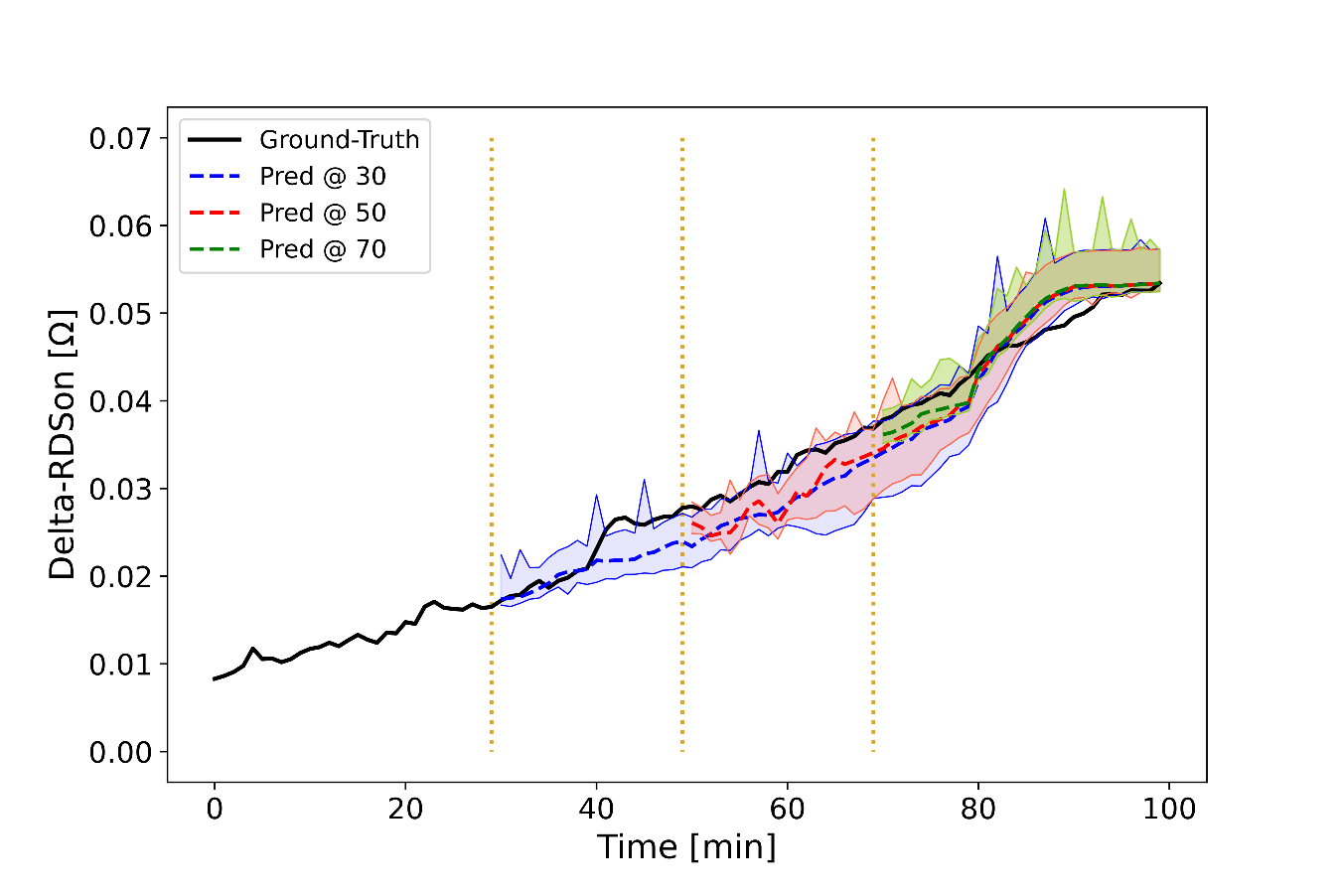}\\
        \vspace{-0.5cm}
 	\includegraphics[width=0.6\textwidth]{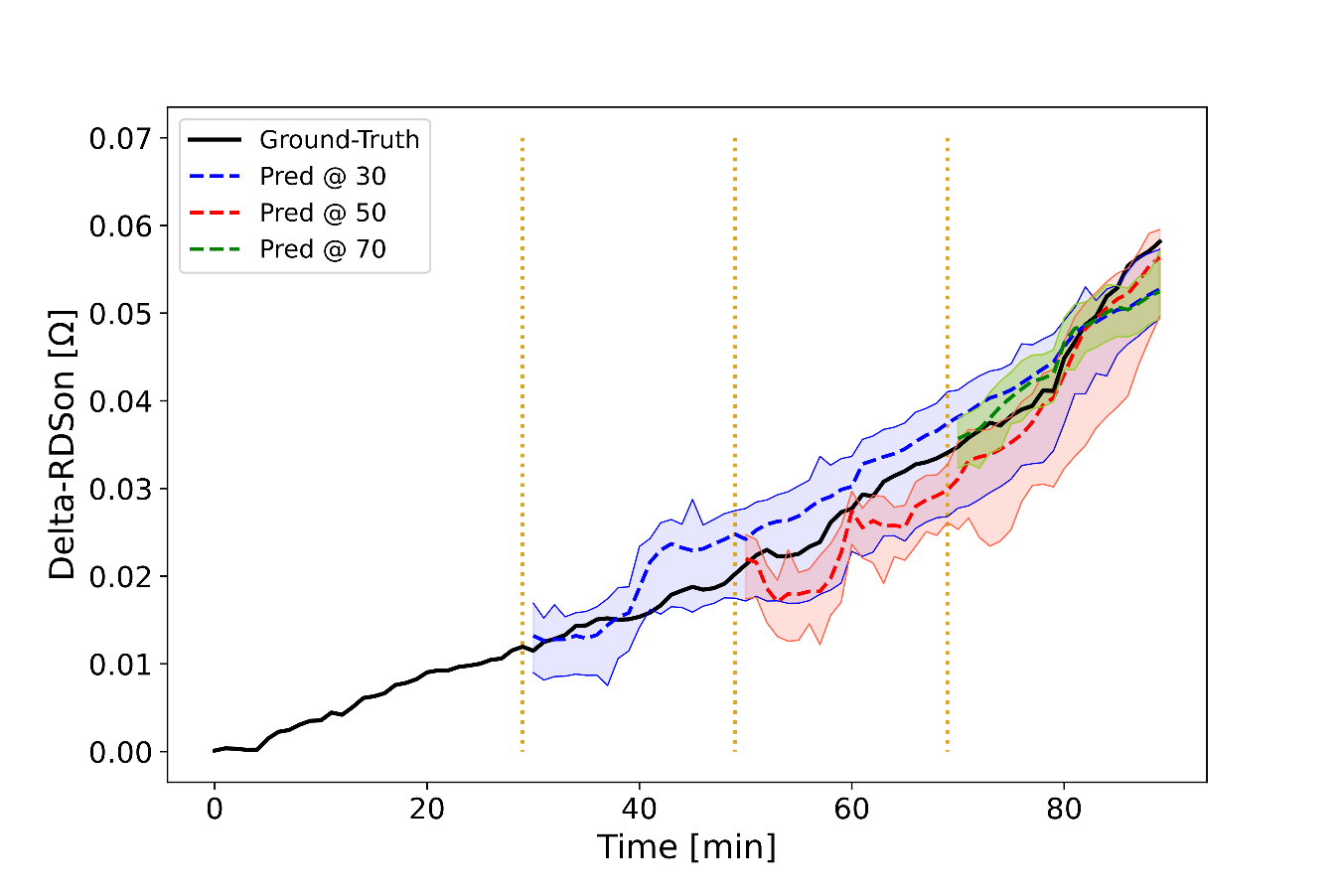}
	\caption{TFT forecasting results (mean and 90\% prediction intervals) for tests 11 (top), 36 (middle) and 9 (bottom). Vertical dashed lines indicate forecasting instants using 30\%, 50\% and 70\% of training data.}
	\label{fig:TFT_long_Test_All}
\end{figure}

Finally, taking the best TFT architecture Figure~\ref{fig:TFT_long_Test_All} shows prediction results with the 90\% prediction intervals for the different tests and prediction horizons.

\textcolor{blue}{Figure~\ref{fig:TFT_long_Test_All} shows that the most accurate prediction results are obtained with the test 36 for all the tested prediction horizons. Furthermore, it can be observed that the prediction intervals are wider for tests 11 and 9, which is in agreement with the accuracy of the obtained results.}

\subsubsection{Attention mechanism}
\label{ss:Attention}

\begin{figure*}[!htb]
\centering
\captionsetup{farskip=0pt,nearskip=4pt}
\subfloat[Attention curve - T36, 70 steps ahead]{\includegraphics[width=0.33\textwidth]{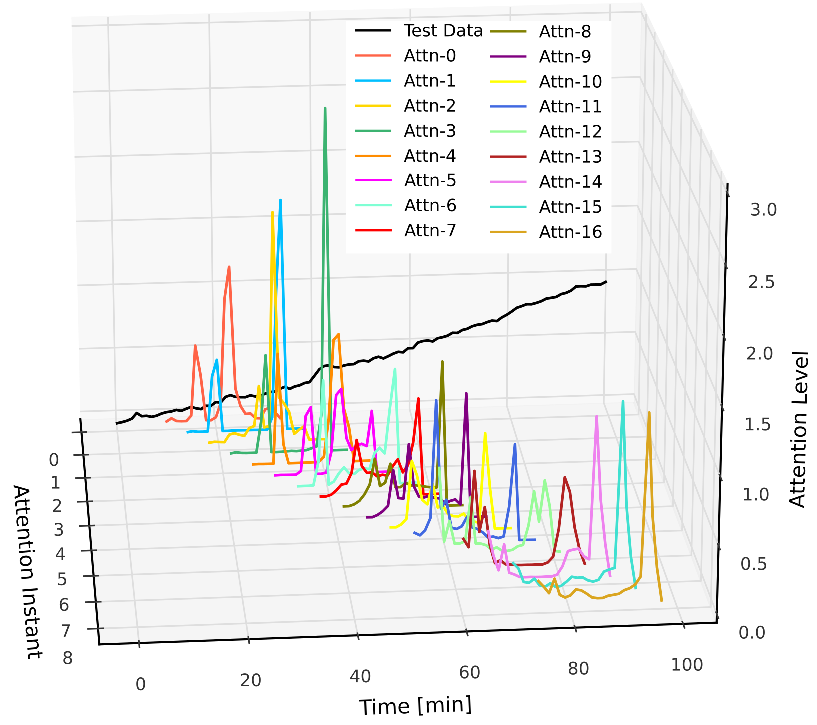}}%
\subfloat[Attention curve - T11, 70 steps ahead]{\includegraphics[width=0.33\textwidth]{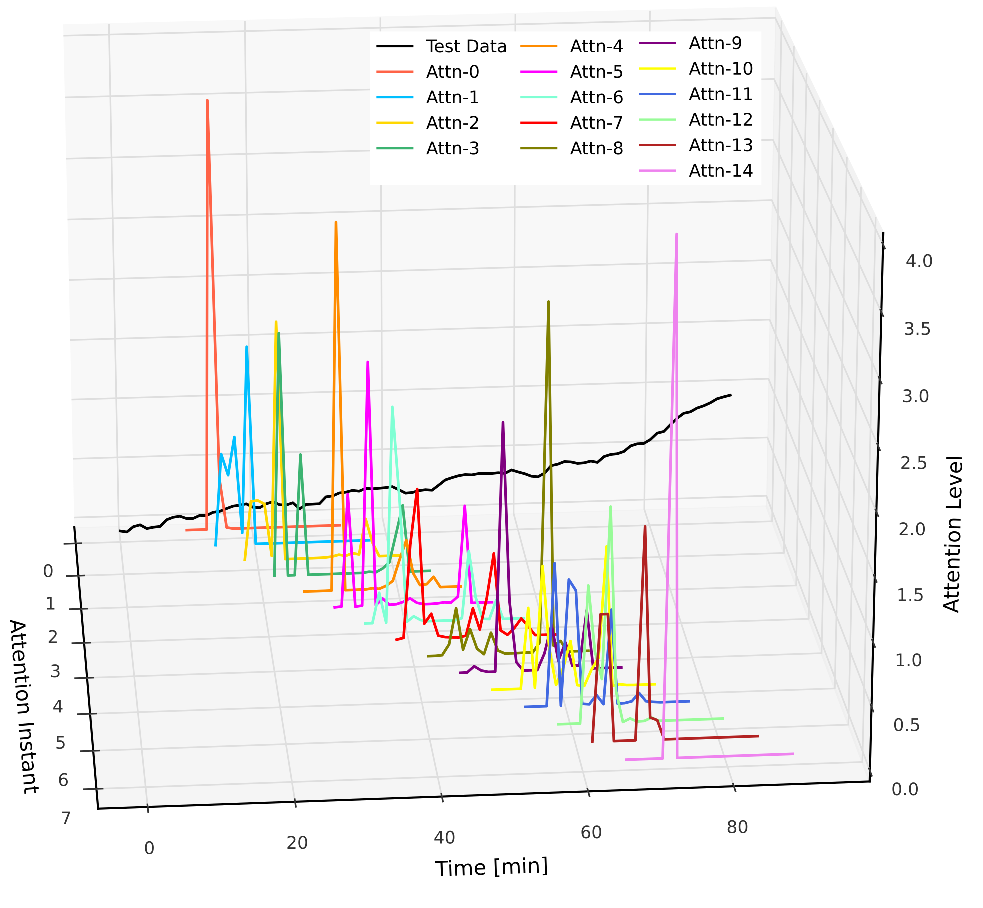}}%
\subfloat[Attention curve - T9, 70 steps ahead]{\includegraphics[width=0.33\textwidth]{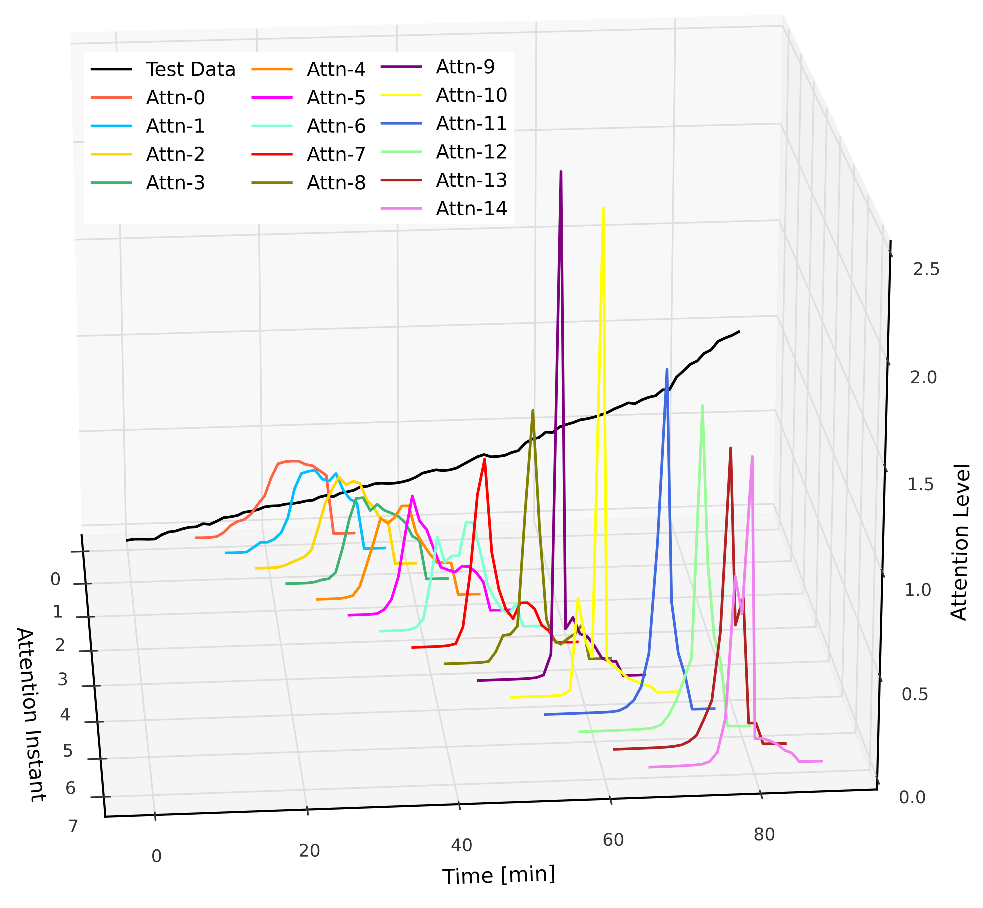}} \\
\subfloat[Max. Attention - T36, 70 steps ahead]{\includegraphics[width=0.33\textwidth]{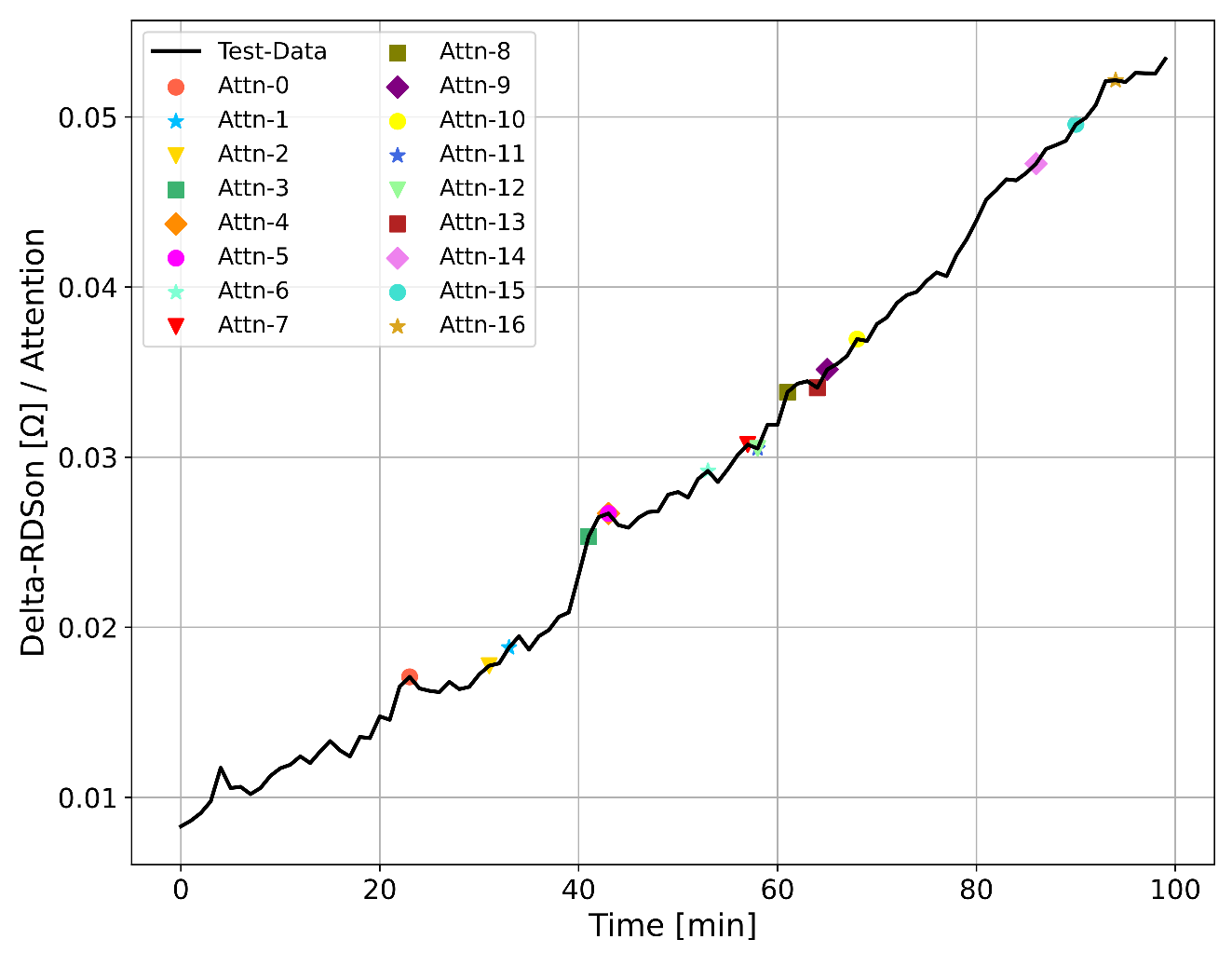}}
\subfloat[Max. Attention - T11, 70 steps ahead]{\includegraphics[width=0.33\textwidth]{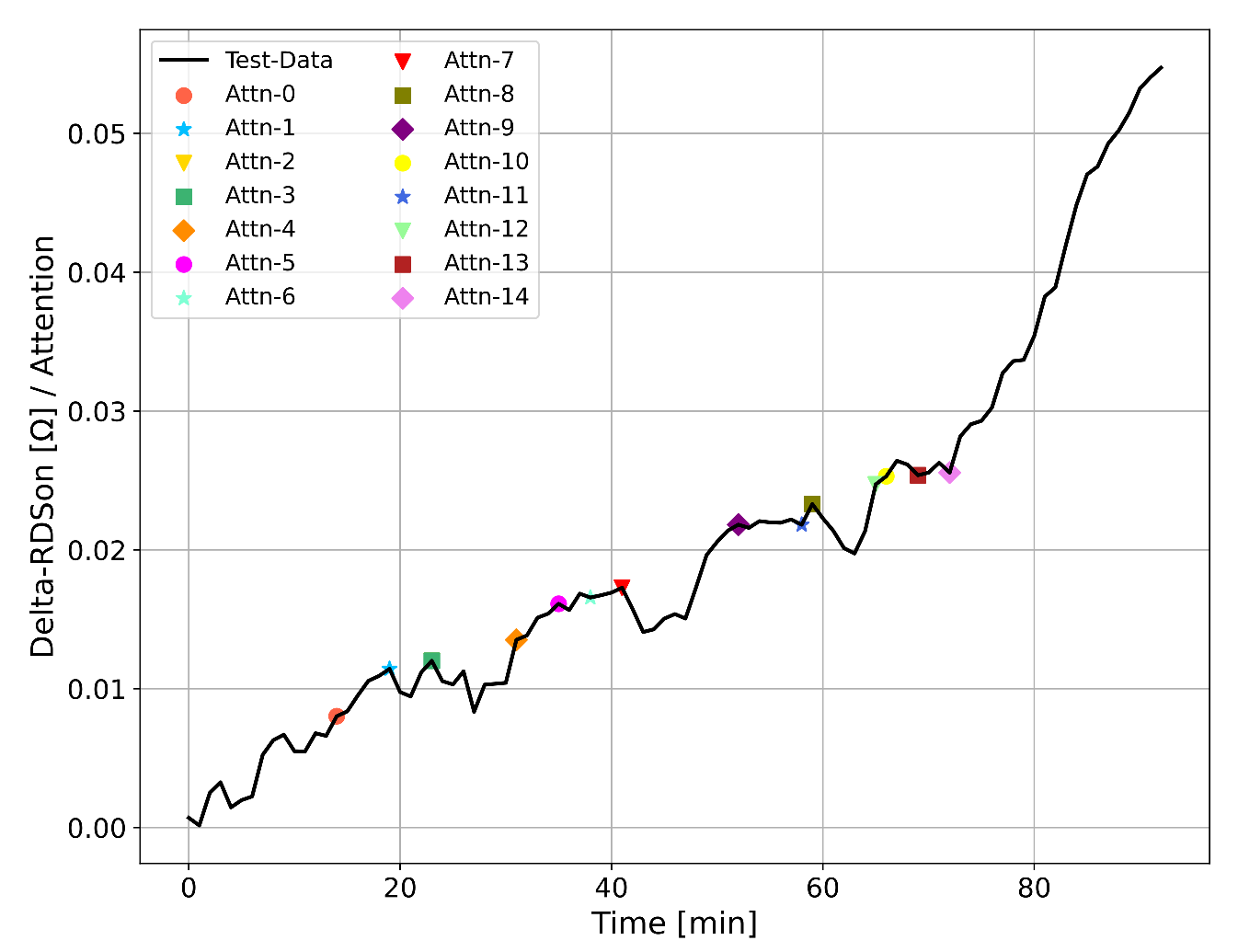}}
\subfloat[Max. Attention - T9, 70 steps ahead]{\includegraphics[width=0.33\textwidth]{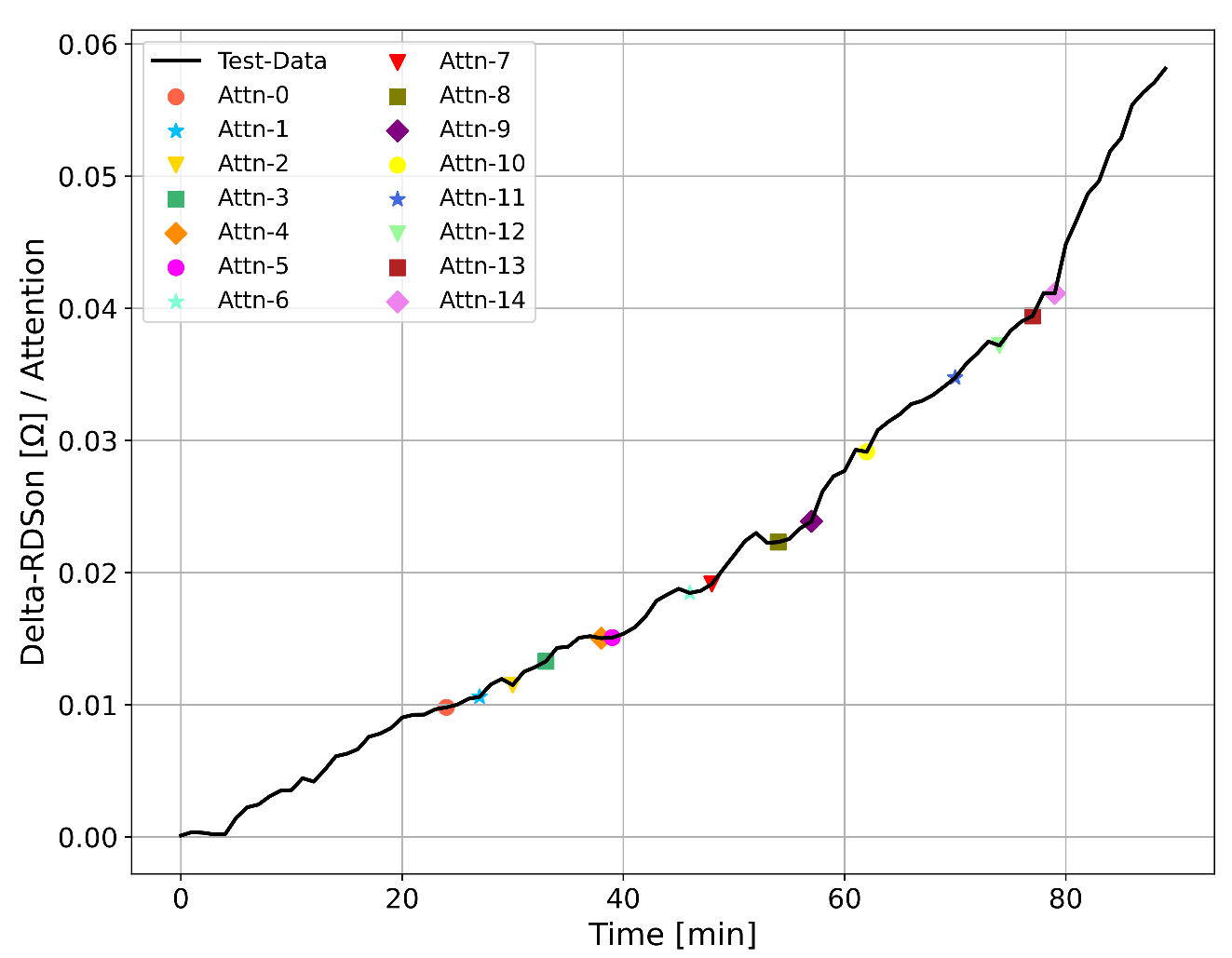}}
\caption{TFT attention plots for different tests. Top: Attention curves. Bottom: maximum attention points with respect to $R_{DS_{ON}}$.}
\label{fig:attCurve_All}
\end{figure*}

The transformer architecture enables the identification of attention points, which are indicators of the input sequences of the TFT architecture that are used to forecast (\citealp{NIPS2017_Vaswani}). Focusing on 70-step ahead predictions (\textit{i.e.}, 30\% of training data), Figures~\ref{fig:attCurve_All} (a)-(c) show the attention curves for the different tests. Note that 70-step-ahead predictions are performed through concatenation. That is, 4 step-ahead predictions are used in an iterative manner to reach 70 step-ahead predictions.

Taking the maximum attention levels in Figures~\ref{fig:attCurve_All}(a)-(c) with their corresponding time instant, Figures~\ref{fig:attCurve_All}(d)-(f) show the identified attention points in the $R_{DS_{ON}}$ signal, which elucidate points where new failure mechanism arise or describe point of exacerbation. It can be observed that the TFT model pays attention to the main signal turning points for each test. In addition, it can be observed that, in general, as the prediction horizon progresses, the attention points become narrower and more focused.

\section{\textcolor{blue}{Discussion}}
\label{sec:Discussion}

\textcolor{blue}{The final objective of the proposed comparative forecasting approach is to examine long- and short- term MOSFET ageing forecasting ability of classical state-space, statistical, and NN-based architectures, and novel transformer-architecture based forecasting models. This will enable to systematically select a prognostics model adapted to different forecasting needs. Yet, the selection of the final prognostics architecture is dependent on different factors and this section aims to discuss relevant criteria for an effective MOSFET ageing forecasting implementations.}

\subsection{\textcolor{blue}{Accuracy and computational cost}}

\textcolor{blue}{Short-term ageing forecasting results (cf. Table~\ref{table:Prognostics_Results} and Figure~\ref{fig:RunToFailure_Tests}) have shown that using classical state-space, statistical and NN-based architectures can be used to predict MOSFET ageing trajectory with E-ELM and ARIMA models obtaining the best results, with a good performance of the UKF model. The TFT model performs with an average accuracy for short-term forecasts, and therefore, for short-term forecasting the use of classical methods is suggested due to the simplicity with respect to the TFT model.}

\textcolor{blue}{As for long-term ageing forecasts (cf. Figure~\ref{fig:tftvselmlong}), the TFT outperforms the best classical forecasting model (E-ELM). It is observed that the designed covariate informs the TFT model and improves prediction accuracy. Namely, the main strength of the TFT architecture is the capability to integrate additional covariates and learn influential dynamics that can be used to predict future ageing values. In contrast, classical state-space and statistical forecasting methods tend to accumulate errors for longer prediction horizons. Similarly, classical NN-based forecasting methods struggle to extrapolate beyond training data, and therefore increase the prediction error for long-term forecasting horizons. Therefore, TFT can be used as an offline approach to elucidate different failure mechanisms and evaluate long-term ageing trajectories.}

\textcolor{blue}{The main differences in performance and computational cost are caused by the architectural complexity of the employed forecasting methods. Classical statistical forecasting methods, rely on autoregressive, moving average, and exponential smoothing techniques to track trends and correct errors. These methods are computationally efficient due to their parametric structure, but may struggle with capturing complex nonlinear degradation patterns.}

\textcolor{blue}{State-space models require the specification of an underlying system model to track the data and update state estimates in real time. Their computational complexity increases with the dimensionality of the state-space and the nonlinearity of the system equations. Traditional data-driven forecasting methods rely on nonlinear transformations to approximate complex degradation trends. Their computational cost scales with the size of the network, the length of historical data sequences, and the training process.}

\textcolor{blue}{The TFT represents the most computationally demanding approach due to its self-attention mechanisms, gating layers, and multi-horizon forecasting architecture. Unlike other methods, the TFT can selectively focus on relevant past information through the attention mechanism, incorporate multi-source inputs through covariates, and provide uncertainty-aware long-term forecasts. The added complexity translates into higher computational cost, especially during training, but as observed in the obtained results, this leads to a superior accuracy when forecasting highly nonlinear and long-term degradation patterns. Table~\ref{table:Summary_Final} displays a qualitative summary of the analysed methods, including their strengths, weaknesses and best application contexts.}

\begin{table}[htb]
\color{blue}
    \centering
    \caption{Qualitative comparison of the analysed methods.}
    \setlength{\tabcolsep}{4.7pt}
        \begin{tabular}{m{4em} m{18em} m{16em} m{8em} }
             \hline
             \textbf{Method} & \textbf{Strengths} & \textbf{Weaknesses} & \textbf{Best Use} \\              \hline
             \textbf{ARIMA, Holt} & Fast, simple, and interpretable. Good for linear trends. Low computational cost. & Difficulties with nonlinear trends & Short-term  \\ 
             \textbf{EKF, UKF} & Robust to nonlinearities. Real-time tracking & Sensitive to state-space. Moderate computational cost & Short-term  \\ 
\textbf{E-NN, E-ELM} & Fast inference. Capture nonlinear and complex trends. & Requires large datasets. Moderate computational cost & Short-, Mid-term \\               
\textbf{TFT} & Good long term accuracy. Handles long-term dependencies & Requires large datasets. Very high computational cost. & Long-term forecasting \\               \hline
        \end{tabular}
        \label{table:Summary_Final}
\end{table}	

\subsection{\textcolor{blue}{TFT for Ageing Forecasting}}

\textcolor{blue}{The TFT architecture has the ability to incorporate covariates along with attention mechanisms that can be very useful for long-term prognostics. However, the adaptation of the TFT architecture for prognostics tasks is not trivial. This is mainly because relevant ad-hoc covariates need to be developed, which inform the TFT architecture about the evolution of the analysed ageing process.}

\subsection{\textcolor{blue}{MOSFET failure mechanisms}}

\textcolor{blue}{While the proposed approach has considered the exponential crack growth as the primary failure mechanism, there are other failure mechanisms present in the analysed ageing process, such as thermal runaway.}

\textcolor{blue}{The impact of additional failure mechanisms becomes more significant beyond a certain key ageing threshold, making it challenging to track the ageing trajectory. This threshold can guide the selection of models. For well-defined modeling equations, such as exponential growth, where additional failure mechanisms emerge in later stages, statistical forecasting and tracking methods may suffice to forecast the ageing trajectory. Conversely, equations describing the failure physics that are ill-defined due to the interaction of multiple nonlinear failure mechanisms can benefit from data-driven models.}

\section{Conclusions}
\label{sec:Conclusions}

MOSFETs are key semiconductor-based power elements for various power-electronic systems. \textcolor{blue}{Their most critical failure mode is bond wire lift-off, which is caused by crack growth due to thermal fatigue. Focusing on this ageing mechanism, this research has evaluated the use of novel and classical forecasting methods for MOSFET ageing forecasting. Namely, Temporal Fusion Transformers (TFTs) have been adapted, implemented and tested for short- and long-term aging forecasting activities.}

TFT models have been designed with ad-hoc covariates, and they have been compared with classical \textcolor{blue}{state-space} tracking methods, including Extended and Unscented Kalman filters (EKF, UKF), classical forecasting methods, such as ARIMA and Holt, and other machine learning models used for forecasting, including ensemble models of Neural Networks (E-NN) and Extreme Learning Machines (ELM).

Obtained results show that, for short-term ageing forecasting horizons, the use of ARIMA and E-ELM models is the most appropriate. On the contrary, for long-term ageing prediction horizons, the use of TFT models greatly improves prediction accuracy with respect to all the tested models. \textcolor{blue}{This highlights the capability of TFT models to incorporate (i) previous information for long-term future predictions making use of the attention mechanism and (ii) ageing-influencing data through covariates.}

\textcolor{blue}{The bond wire lift-off phenomena manifests as a exponential crack growth}. The proposed aging forecasting approach allows obtaining an accurate modeling, which is very appropriate for short-term ageing prediction. With this model, all the analysed classic forecasting algorithms (EKF, UKF, ARIMA, Holt, E-ELM, E-NN) result in a proper short-term ageing estimation. \textcolor{blue}{Given the relatively large number of computations of the TFT model, classic forecasting algorithms may be used for on-line prediction.}

For long-term prediction, classic forecasting algorithms are unable to obtain accurate results because, as the ageing of the MOSFET progresses, different failure modes interact and this is not captured. \textcolor{blue}{In this context, TFTs represent an appropriate ageing forecasting alternative as they can incorporate multiple sources of information through covariates and focus on the most relevant information through the attention mechanism. To benefit from the TFT architecture for ageing forecasting, it is necessary to design and incorporate ageing covariates,} which inform about the expected future inputs in the form of failure progression. 

The obtained results have shown that TFTs can obtain accurate forecasting results for long-term ageing forecasting. In addition, TFT attention points identify key ageing turning points, which are indicative of new failure modes or excessive accelerated ageing. All in all, it can be concluded that TFTs can be used to build MOSFET prognostics models. However, their computational cost may be an implementation limit, compared with classical algorithms, especially for on-line condition monitoring applications.

\textcolor{blue}{Future work could explore computationally efficient long-term ageing forecasting models, opening new opportunities in edge machine learning  (\citealp{hua2023edge}). This could enable low-latency, on-device inference in online monitoring systems for asset management (\citealp{Garro2019}). Another interesting direction is the adaptation of foundation models (\citealp{Yuxuan_24}) for MOSFET ageing prediction. These large-scale pre-trained models could be fine-tuned for long-term degradation forecasting, potentially improving generalization across different operational conditions and component types}.

\section*{Code Availability}

The code will be \textcolor{blue}{available on the website: https://github.com/joxeina/AgeingForecastingMOSFETs}.

\section*{Acknowledgements}

This research was partially funded by the Basque Government, Department of Education (grant No. KK-2024/00030 and IT1504-22). J. I. Aizpurua is funded by the Ramón y Cajal Fellowship, Spanish State Research Agency (grant number RYC2022-037300-I), co-funded by MCIU/AEI/10.13039/501100011033 and FSE+.

\bibliographystyle{unsrtnat}
\bibliography{references}

\end{document}